\documentclass[11pt,draftcls, onecolumn]{IEEEtran}
\usepackage{epsfig}
\usepackage{afterpage}
\usepackage{graphicx}
\usepackage{amsmath}
\usepackage{amsfonts}
\usepackage{amssymb}
\usepackage{cite}
\usepackage{psfrag}
\usepackage{subfigure}

\ifCLASSINFOpdf
\else
\fi
\begin{document}
%
\title{Joint Filter Design of Alternate MIMO AF Relaying Networks with Interference Alignment}

%
%

\author{Ki-Hong~Park,~\IEEEmembership{Member,~IEEE,}
        and~Mohamed-Slim~Alouini,~\IEEEmembership{Fellow,~IEEE,}
\thanks{K.-H. Park and M.-S. Alouini are with the Electrical Engineering Program, Physical Science and
Engineering Division, King Abdullah University of Science and Technology (KAUST), Thuwal, Makkah Province,
Kingdom of Saudi Arabia (Email: \{kihong.park; slim.alouini\}@kaust.edu.sa).}
\thanks{This paper is an extended version of work accepted for presentation at the {\it IEEE International Conference on Communications}
(ICC'12), Ottawa, Canada, June 2012.
}
}

\IEEEpubid{0000--0000/00\$00.00~\copyright~2007 IEEE}


\maketitle

\vspace{-0.2in}
\begin{abstract}
We study in this paper a two-hop relaying network consisting of one source, one destination, and three amplify-and-forward (AF) relays operating in a half-duplex mode. In order to compensate for the inherent loss of capacity pre-log factor $\frac{1}{2}$ in a half-duplex mode, we consider alternate transmission protocol among three relays where two relays and the other relay alternately forward messages from source to destination. We consider a multiple-antenna environment where all nodes have $M$ antennas. Aligning the inter-relay interference due to the alternate transmission is utilized to make additional degrees of freedom (DOFs) and recover the pre-log factor loss. It is shown that the proposed relaying scheme can achieve $\frac{3M}{4}$ DOFs compared with the $\frac{M}{2}$ DOFs of conventional AF relaying. In addition, suboptimal linear filter designs for a source and three relays are proposed to maximize the system achievable sum-rate for different fading scenarios when the destination utilizes a linear minimum mean-square error filter for decoding. We verify from our selected numerical results that the proposed filter designs give significant improvement over a naive filter or conventional relaying schemes.
\end{abstract}


%
\IEEEpeerreviewmaketitle

\section{Introduction}
%
%
%
%
\IEEEPARstart{W}{ireless} relaying systems have been recently considered an attractive option because of their potentials to improve the system throughput, enhance the cell-edge performance, extend cell coverage, and reduce the overall system deployment cost~\cite{Wu1,Pabst1,Nostratinia1,Sendonaris1,Sendonaris2,Laneman1,Boyer1}. As such they have been considered for the standardization of IEEE 802.16j, 16m and 3GPP LTE-Advanced~\cite{Sydir1,ieee16j,ieee16m,lte_adv}. Most of the nodes in a relaying system operate in a half-duplex mode, which means that they cannot transmit and receive the signal simultaneously and act as a transmitter or receiver at the same time. This inherent structural property of a half-duplex system requires two time slots in two-hop relaying networks since the source transmits the signal to relays during the first time slot/phase and the relays forward the received signal to the destination during the second time slot/phase.
It results in a loss of capacity pre-log factor of $\frac{1}{2}$ in the half-duplex protocol\footnote{In general, the capacity pre-log factor is also referred to as the degree of freedom (DOF).}.


There has been a steady interest in order to overcome the inherent disadvantage of half-duplex relaying systems. First, the incremental relaying protocol has been proposed in \cite{Laneman1}. The source broadcasts the message first and the relay is only used to retransmits the message from the source in an attempt to exploit spatial diversity just in case that the destination fails to decode the message from the received signal through a direct link between source and destination. For the non-orthogonal amplify-and-forward (NAF) protocol~\cite{Nabar1,Krikidis1}, the source transmits a new message to the destination during the second time slot. This cooperative relaying system during two time slots is equivalently modeled as a multiple-input multiple-output (MIMO) system which can compensate for the loss of capacity pre-log factor in a half-duplex mode~\cite{Nabar1}. The aforementioned methods are utilized in the relaying systems assuming that direct transmission from source to destination is available. In the absence of a direct link between source and destination due to a deep fade or block by obstacles, two-way relaying and two-path relaying have been proposed~\cite{Rankov1}. In the two-way relaying protocol, the bidirectional connection between source and destination is established to compensate for the loss in capacity pre-log factor.

On the other hand, the two-path relaying protocol adjusts the phase difference where the source alternately transmits the signals to the destination via different relays. One relay receives the signal from the source while the other relay forwards the message to the destination. In this protocol, the desired signal forwarded to the destination acts as an inter-relay interference to the relay in receiver mode. In \cite{Rankov1}, the destination utilizes successive decoding with successive interference cancelation and the proposed method gives good performance improvement only for a weak to moderate inter-relay channel. The authors in \cite{wicaksana09} proposed canceling the inter-relay self interference at one of the relays and highlighted that its method is still robust even in a strong inter-relay channel. The previous works for two-path relaying in~\cite{Rankov1,wicaksana09} focused on a single-antenna environment and have been extended to a multiple-antenna scenario in~\cite{Boccardi08,parkhs08}. The work in \cite{Boccardi08} exploits two relays with multiple antennas to cancel the inter-relay interference when two relays perform alternate relaying, while this method cannot recover a loss of capacity pre-log factor and only improve the signal-to-noise ratio (SNR) gain. Even though the proposed schemes in \cite{parkhs08} enhance the capacity pre-log factor of the proposed schemes, inter-relay interference is not considered thoroughly by assuming that it is blocked by large obstacles.

In~\cite{psh11}, we simply investigated a decode-and-forward (DF) alternate relaying system with three relays. The proposed scheme was shown to partially compensate for a loss of DOFs by aligning the inter-relay interference from different nodes and making additional spatial dimensions, which has been recently developed for MIMO interference and $X$ networks~\cite{cadambe08,cadambe09}. In this paper, we propose an alternate relaying protocol with source, destination and three amplify-and-forward (AF) relays with multiple antennas in order to compensate for a loss of capacity pre-log factor in case of multiple-antenna scenario. Compared with DF relaying scheme, AF relaying scheme requires much less delay and power consumption since the signal processing and quantizing operation for decoding is unnecessary at the relay. More specifically, in this paper, inter-relay interference alignment (IA) is performed at two relays of all, while source and three relays should participate in the alignment operation in DF relaying. In particular, in the proposed scheme, the IA is embedded to align the inter-relay interference from two relays just in case the two relays forward the message to the destination and the other relay receives the signal from the source. The direct link between source and destination is not considered and it is more troublesome to recover the loss of pre-log factor since the direct link inherently ensures full pre-log factor even without relaying links. We show that the proposed method can achieve $\frac{3M}{4}$ DOFs compared with $\frac{M}{2}$ DOFs of conventional AF relaying when all nodes are equipped with $M$ antennas. Linear filters are considered at the source, relay, and destination side, respectively. We then propose a class of linear filters at source and relays that can maximize the system achievable sum-rate for different fading scenarios. The proposed filter design is based on utilizing the subgradient method consecutively and alternately. We verify that the proposed filters are robust and give significant improvement over a naive filter and conventional relaying schemes though they only guarantee a local maximum of achievable sum-rate. In addition, we propose distributed algorithm to find the amplifying filters at the relays which do not have to mutually exchange the channel information in order to align the inter-relay interference signals though there is a decrease in rate due to a reduction in costs of the interchange of channel information.

The remainder of this paper is organized as follows. In section~\ref{sec_sys}, we introduce the system model of an alternate AF relaying protocol with three relays. Section~\ref{sec_design} describes the source and relay filter designs for different fading scenarios. We present our numerical examples in section~\ref{sec_sim} and a brief conclusion summarizing the main results and discussing future works of the paper are given in section~\ref{sec_con}.

Throughout the paper, we use the following notations. Upper and lower case boldfaces are used to employ matrices and vectors, respectively. $\mathbf{I}_m$ denotes an $m\times m$ identity matrix.
$\mathbf{A}^{\mathsf{T}}$, $\mathbf{A}^{\mathsf{*}}$, $\mathbf{A}^{\mathsf{H}}$, and $\mathbf{A}^{\mathsf{-1}}$ denote the transpose, conjugate, Hermitian transpose, and the inverse of an arbitrary matrix $\mathbf{A}$, respectively. $\mathbf{A}_{b:c}$ denotes a submatrix consisting of the $a$th to $b$th column vectors of matrix $\mathbf{A}$. $\mathrm{Span}(\mathbf{A})$ represents the space spanned by matrix $\mathbf{A}$ and $\mathrm{Span}(\mathbf{A})\perp \mathrm{Span}(\mathbf{B})$ means that the vector spaces of matrices, $\mathbf{A}$ and $\mathbf{B}$, are orthogonal. $\mathrm{tr}\{\cdot\}$, $\mathbb{E}\!\left[\cdot\right]$, and $Re(\cdot)$ denote the trace, expectation, and real part of complex scalar, vector, and matrix, respectively.


\section{System model}
\label{sec_sys}

In this paper, we consider a half-duplex relay network consisting of one source, one destination, and three AF relays which are denoted as $S$, $D$ and $R_i$ for $i\in\{1,2,3\}$, respectively. Each node is equipped with $M$ even antennas and cannot transmit and receive data simultaneously in a half-duplex mode. We assume that the channel between two nodes is block fading during transmission and a channel matrix from the $j$th node to the $i$th node for the $n$th time slot is defined as $\mathbf{H}_{ij}[n]\in\mathbb{C}^{M\times M}$ for $i, j\in\{S,D,1,2,3\}$ and $i\neq j$. We also assume that the direct link between $S$ and $D$ is negligible due to a large path loss.

\begin{figure}[tb]
\epsfxsize 5in \centerline{\hbox{\epsffile{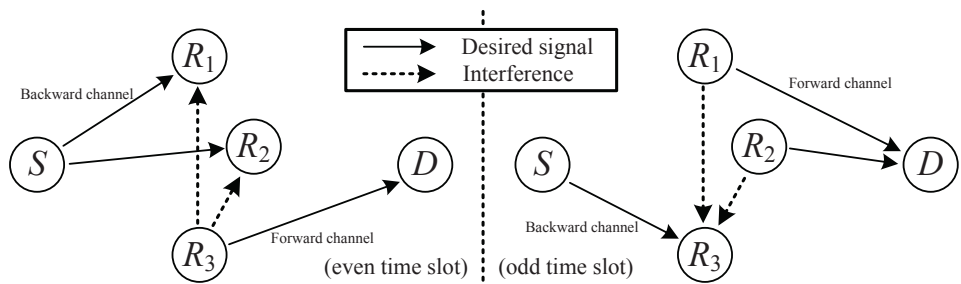}}}
\caption{The proposed dual-hop half-duplex protocol.}\label{fig1}
\end{figure}

In Fig.~\ref{fig1}, we illustrate the system model of our proposed method for successive two time slots. At each time slot, $S$ sends transmit signals to the relays and the other relays forward the received signals to $D$ in an alternate way. This transmission protocol is consecutively repeated every two time slots as summarized in Table~\ref{protocol}. For the even time slots, $S$ transmits $M$ data streams to $R_1$ and $R_2$, while $R_3$ forwards the received signal at the previous time slot to $D$. At the odd time slots, $S$ sends $\frac{M}{2}$ data streams to $R_3$, while $R_1$ and $R_2$ forward the signal received at the previous time slot to $D$. It is equivalently viewed as $2\times 3$ or $3\times 2$ interference $Z$ channels and optimal in terms of achievable DOFs.

The symbol vector, $\mathbf{s}[n]$, at $S$ is generated from an independently encoded Gaussian codebook with
$\mathbf{s}[n]\sim\mathcal{N}(\mathbf{0},\mathbf{I}_{\!M})$ for even $n$, and $\mathbf{s}[n]\sim\mathcal{N}(\mathbf{0},\mathbf{I}_{\!\frac{M}{2}})$ for odd $n$. The symbol vector is beamformed by a linear precoding filter matrix, $\mathbf{T}[n]\in\mathbb{C}^{M\times M}$ for even $n$ and $\mathbf{T}[n]\in\mathbb{C}^{M\times \frac{M}{2}}$ for odd $n$. Then, the transmit signal vector of the $n$th time slot can be written as
$\mathbf{x}[n]=\sqrt{p_t[n]}\mathbf{T}[n]\mathbf{s}[n]\in\mathbb{C}^{M}$.
We assume that the total transmit power for each transmission at $S$ is limited to $P_S$, which is given by
$\mathrm{tr}\left\{\mathbb{E}\!\left[\mathbf{x}[n]\mathbf{x}^{\mathsf{H}}[n]\right]\right\}=\mathrm{tr}\left\{p_t[n]\mathbf{T}[n]\mathbf{T}^{\mathsf{H}}[n]\right\}= P_S$, where $p_t[n]$ is a normalization factor to satisfy total power constraint.
Then, the received signal at the relay side for each time slot is defined as
\begin{eqnarray}
\begin{cases}
\mathbf{y}_i[n]=\mathbf{H}_{i\!S}[n]\mathbf{x}[n]+\mathbf{H}_{i3}[n]\mathbf{x}_3[n]+\mathbf{z}_i[n],  &\textrm{even }n\;(i=1,2) \\
\displaystyle\mathbf{y}_3[n]=\mathbf{H}_{3\!S}[n]\mathbf{x}[n]+\sum_{i=1}^2\mathbf{H}_{3i}[n]\mathbf{x}_i[n]+\mathbf{z}_3[n], &\textrm{odd }n,\\
\end{cases}\label{y_relay}
\end{eqnarray}
where $\mathbf{x}_i[n]\in\mathbb{C}^{M}$ is the transmit signal of the $i$th relay amplified and forwarded to the destination and $\mathbf{z}_i[n]\in\mathbb{C}^{M}$ is a complex white Gaussian noise vector with $\mathcal{CN}(\mathbf{0},\sigma_i^2\mathbf{I}_{\!M})$. At each relay, the received signal is multiplied by an amplifying matrix, $\mathbf{F}_i[n]\in\mathbb{C}^{M\times M}$, and the transmit signal is computed as
\begin{eqnarray}
\mathbf{x}_i[n]=\sqrt{p_i[n]}\mathbf{F}_i[n]\mathbf{y}_i[n\!-\!1],\label{tx_relay}
\end{eqnarray}
where $p_i[n]$ is a normalization scalar factor of the $i$th relay to satisfy the total power constraint.
We assume that the total transmit power at each relay node is constrained on a certain power, $P_R$, which is given by $\mathrm{tr}\left\{\mathbb{E}\!\left[\mathbf{x}_i[n]\mathbf{x}_i^\mathsf{H}[n]\right]\right\}=\mathrm{tr}
\left\{p_i[n]\mathbf{F}_i[n]\boldsymbol\Sigma_i[n\!-\!1]\mathbf{F}_i^\mathsf{H}[n]\right\}=P_R$,
where we define $\boldsymbol\Sigma_i[n]=\mathbb{E}\!\left[\mathbf{y}_i[n]\mathbf{y}_i^\mathsf{H}[n]\right]$.

In (\ref{y_relay}), the second terms of the received signal, $\mathbf{H}_{i3}[n]\mathbf{x}_3[n]$ and $\sum_{i=1}^2\mathbf{H}_{3i}[n]\mathbf{x}_i[n]$, are referred to as the inter-relay interference from other relays. We focus on perfectly canceling the inter-relay interference\footnote{These interference signals consisting of previous signals degrade the performance of current received signal at $D$ and make the implementation of the relay and destination side complicated in order to alleviate the effect of the interference and detect the desired data. In addition, since the inter-relay interference signal includes causal channel knowledge for all the previous time slot, it requires to use a large amount of memories at the relay and destination sides.} when exploiting an amplifying matrix as in (\ref{tx_relay}) before forwarding the transmit signal at the relay side. In order to ensure zero inter-relay interference for each time slot, it is required that
\begin{eqnarray}
\begin{cases}
\displaystyle\mathbf{F}_3[n]\sum_{i=1}^2\mathbf{H}_{3i}[n\!-\!1]\mathbf{F}_i[n\!-\!1]\mathbf{y}_i[n\!-\!2]=\mathbf{0}, &\textrm{even }n \\
\mathbf{F}_i[n]\mathbf{H}_{i3}[n\!-\!1]\mathbf{F}_3[n\!-\!1]\mathbf{y}_3[n\!-\!2]=\mathbf{0}, &\textrm{odd }n \;(i=1,2). \\
\end{cases}\label{ziri_const}
\end{eqnarray}
Under the constraints in (\ref{ziri_const}), we note that the covariance matrix of the received signal at the $i$th relay, $\mathbf{y}_i[n]$, can be rewritten as $\boldsymbol\Sigma_i[n]=p_t[n]\mathbf{H}_{i\!S}[n]\mathbf{T}[n]\mathbf{T}^{\mathsf{H}}[n]
\mathbf{H}_{i\!S}^{\mathsf{H}}[n]+\sigma_i^2\mathbf{I}_{\!M}$.
The transmit signal at the relay side is forwarded to the destination for each time slot and the received signal at $D$ is finally written as
\begin{eqnarray}
\mathbf{y}_{\!D}[n]=\sqrt{p_t[n\!-\!1]}\tilde{\mathbf{H}}[n,n\!-\!1]\mathbf{s}[n\!-\!1]+
\tilde{\mathbf{z}}_{\!D}[n,n\!-\!1],\label{y_d}
\end{eqnarray}
where $\tilde{\mathbf{H}}[n,n\!-\!1]$ is the effective channel matrix of the $n\!-\!1$th data symbol vector for the $n\!-\!1$th to $n$th time slot, which is defined as
\begin{eqnarray}
\tilde{\mathbf{H}}[n,n\!-\!1]=
\begin{cases}
\sqrt{p_3[n]}\mathbf{H}_{\!D3}[n]\mathbf{F}_3[n]\mathbf{H}_{3\!S}[n\!-\!1]\mathbf{T}[n\!-\!1], & \textrm{even }n \\
\displaystyle\sum_{i=1}^2\sqrt{p_i[n]}\mathbf{H}_{\!Di}\mathbf{F}_i[n]\mathbf{H}_{i\!S}[n\!-\!1]\mathbf{T}[n\!-\!1], & \textrm{odd }n,\\
\end{cases}\nonumber
\end{eqnarray}
and $\tilde{\mathbf{z}}_{\!D}[n,n\!-\!1]$ is the Gaussian noise which is defined as
\begin{eqnarray}
\tilde{\mathbf{z}}_{\!D}[n,n\!-\!1]=
\begin{cases}
\sqrt{p_3[n]}\mathbf{H}_{\!D3}[n]\mathbf{F}_3[n]\mathbf{z}_3[n\!-\!1]+\mathbf{z}_{\!D}[n], &\textrm{even }n \\
\displaystyle \sum_{i=1}^2\sqrt{p_i[n]}\mathbf{H}_{\!Di}[n]\mathbf{F}_i[n]\mathbf{z}_i[n\!-\!1]+\mathbf{z}_{\!D}[n], &\textrm{odd }n. \\
\end{cases} \nonumber
\end{eqnarray}
$\mathbf{z}_{\!D}[n]\in\mathbb{C}^M$ is a complex white Gaussian noise vector with $\mathcal{CN}(\mathbf{0},\sigma_{\!D}^2\mathbf{I}_{\!M})$ at $D$.
We consider a linear filter, $\mathbf{W}_{\!D}[n]$, at $D$ and the estimated data symbol vector is given by $\hat{\mathbf{s}}[n\!-\!1]=\mathbf{W}_{\!D}^{\mathsf{H}}[n]\mathbf{y}_{\!D}[n]$.
The mean square error (MSE) matrix of the $n\!-\!1$th data vector at $D$ can be computed by
\begin{eqnarray}
\mathbf{E}[n,n\!-\!1]\!\!&\!\!=\!\!&\!\!\mathbb{E}\!\left[\left(\hat{\mathbf{s}}[n\!-\!1]-\mathbf{s}[n\!-\!1]\right)\left(\hat{\mathbf{s}}[n\!-\!1]-\mathbf{s}[n\!-\!1]\right)^{\mathsf{H}}\right]\nonumber\\
\!\!&\!\!=\!\!&\!\!\left(\sqrt{p_t[n\!-\!1]}\mathbf{W}_{\!D}^{\mathsf{H}}[n]\tilde{\mathbf{H}}[n,n\!-\!1]-\mathbf{I}[n]\right)\left(\sqrt{p_t[n\!-\!1]}\mathbf{W}_{\!D}^{\mathsf{H}}[n]
\tilde{\mathbf{H}}[n,n\!-\!1]-\mathbf{I}[n]\right)^\mathsf{H}\nonumber\\
\!\!&\!\!+\!\!&\!\!\mathbf{W}_{\!D}^{\mathsf{H}}[n]\boldsymbol\Sigma_{\tilde{\mathbf{z}}}[n]\mathbf{W}_{\!D}[n],\label{mse_matrix}
\end{eqnarray}
where the covariance matrix of $\tilde{\mathbf{z}}[n,n\!-\!1]$ can be calculated as
\begin{eqnarray}
\boldsymbol\Sigma_{\tilde{\mathbf{z}}}[n]\!\!&\!\!=\!\!&\!\!\mathbb{E}\!\left[\tilde{\mathbf{z}}_{\!D}[n,n\!-\!1]\tilde{\mathbf{z}}_{\!D}^{\mathsf{H}}[n,n\!-\!1]\right]
=\begin{cases}
\sigma_3^2p_3[n]\mathbf{H}_{\!D3}[n]\mathbf{F}_3[n]\mathbf{F}_3^{\mathsf{H}}[n]\mathbf{H}_{\!D3}^{\mathsf{H}}[n]+\sigma_{\!D}^2\mathbf{I}_{\!M}, &\textrm{even }n \\
\displaystyle \sum_{i=1}^2 \sigma_i^2p_i[n]\mathbf{H}_{\!Di}[n]\mathbf{F}_i[n]\mathbf{F}_i^{\mathsf{H}}[n]\mathbf{H}_{\!Di}^{\mathsf{H}}[n]+\sigma_{\!D}^2\mathbf{I}_{\!M},
&\textrm{odd }n,\\
\end{cases}\nonumber
\end{eqnarray}
and an identity matrix for each time slot is given by $\mathbf{I}[n]=\mathbf{I}_{\!\frac{M}{2}}$ for even $n$ and $\mathbf{I}[n]=\mathbf{I}_{\!M}$ for odd $n$.
The MSE-optimal linear filter to minimize the MSE matrix is the Wiener filter~\cite{kay93} given in this case by
$\mathbf{W}_{\!D}[n]=\left(p_t[n\!-\!1]\tilde{\mathbf{H}}[n,n\!-\!1]\tilde{\mathbf{H}}^{\mathsf{H}}[n,n\!-\!1]+
\boldsymbol\Sigma_{\tilde{\mathbf{z}}}[n]\right)^{\mathsf{-1}}\!\!\!\!\sqrt{p_t[n\!-\!1]}\tilde{\mathbf{H}}[n,n\!-\!1]$.
Plugging this minimum MSE (MMSE) filter into (\ref{mse_matrix}), the MSE matrix can be rewritten after some manipulations as
\begin{eqnarray}
\mathbf{E}[n,n\!-\!1]=\left(\mathbf{I}[n]+p_t[n\!-\!1]\tilde{\mathbf{H}}^{\mathsf{H}}[n,n\!-\!1]
\boldsymbol\Sigma_{\tilde{\mathbf{z}}}^{\mathsf{-1}}[n]\tilde{\mathbf{H}}[n,n\!-\!1]\right)^{\mathsf{-1}}
.\label{basic_E}
\end{eqnarray}
The achievable sum-rate of the $n\!-\!1$th data vector between $S$ and $D$ can be written as
\begin{eqnarray}
I[n,n\!-\!1]=\log_2 \det \mathbf{E}^{-1}[n,n\!-\!1].
\end{eqnarray}
In order to obtain the above sum-rate, we should find the amplifying matrix filters at the relay side which satisfy the constraint in (\ref{ziri_const}). In the next section, we design a linear filter for each relay to cancel the inter-relay interference. In addition, a linear precoder at $S$ and amplifying filters at $R_i$ are developed to maximize the sum-rate according to different channel assumptions, i.e., slow and fast block fading.

\section{Source/Relay Linear Filter Design}
\label{sec_design}

In order to find the valid amplifying matrix which can perfectly remove the inter-relay interference for each time slot, let us recall the constraints in (\ref{ziri_const}).

For odd time slot, the following two conditions should be met to cancel the inter-relay interference at $R_1$ and $R_2$:
\begin{eqnarray}
\mathrm{Span}\!\left(\mathbf{F}_i[n]\right)\!\!&\!\!\perp\!\!&\!\!\mathrm{Span}\!\left(\mathbf{H}_{i3}[n\!-\!1]
\mathbf{F}_3[n\!-\!1]\right),\;\;\;i=1,2.\label{const1and2}
\end{eqnarray}
Since $\mathbf{F}_3[n\!-\!1]$ is the amplifying matrix which is used to forward the received signal at the previous odd $n\!-\!2$th time slot and $\frac{M}{2}$ symbols are transmitted to $R_3$, we can design a linear amplifying matrix with $\mathrm{rank}(\mathbf{F}_3[n\!-\!1])=\frac{M}{2}$ without loss of DOFs. At this time, if we design a $\mathrm{rank}$-$\frac{M}{2}$ amplifying matrix at $R_3$, we can guarantee the dimensions of inter-relay interference subspace less than $\frac{M}{2}$ ones at $R_1$ and $R_2$, i.e., $\mathrm{rank}\left(\mathbf{H}_{13}[n\!-\!1]
\mathbf{F}_3[n\!-\!1]\right)\leq\frac{M}{2}$ and $\mathrm{rank}\left(\mathbf{H}_{23}[n\!-\!1]
\mathbf{F}_3[n\!-\!1]\right)\leq\frac{M}{2}$. It means that there exists a subspace with the dimensions equal to or more than $\frac{M}{2}$ which is orthogonal to the inter-relay interference subspace for each relay. Therefore, we can design $\mathrm{rank}$-$\frac{M}{2}$ amplifying matrices, $\mathbf{F}_1[n]$ and $\mathbf{F}_2[n]$, for $R_1$ and $R_2$.

Meanwhile, for even time slot, the constraint in (\ref{ziri_const}) can be equivalently rewritten as
\begin{eqnarray}
\mathrm{Span}\!\left(\mathbf{F}_3[n]\right)\perp\bigcup_{i=1}^2\mathrm{Span}\left(\mathbf{H}_{3i}[n\!-\!1]\mathbf{F}_i
[n\!-\!1]\right).\nonumber
\end{eqnarray}
$\mathbf{F}_3[n]$ is required to have at least $\frac{M}{2}$ dimensions to forward the received signal without loss of DOFs. However, the inter-relay interference signal at $R_3$, $\sum_{i=1}^2\mathbf{H}_{3i}[n\!-\!1]\mathbf{F}_i
[n\!-\!1]$, has $M$ dimensions at most without taking into any consideration to reduce its dimensions. We here note that each of amplifying matrices, $\mathbf{F}_1[n\!-\!1]$ and $\mathbf{F}_2[n\!-\!1]$, are designed to have $\frac{M}{2}$ dimensions at the odd time slot. We consider an IA, where the signals can be
designed to cast overlapping shadows at $R_3$, while they remain distinguishable at $D$. Two interference signals can be perfectly aligned on the $\frac{M}{2}$-dimensional subspace if we utilize the amplifying matrices, $\mathbf{F}_1[n\!-\!1]$ and $\mathbf{F}_2[n\!-\!1]$, satisfying the following relation:
\begin{eqnarray}
\mathrm{Span}\left(\mathbf{F}_3[n]\right)\perp\mathrm{Span}\left(\mathbf{H}_{31}[n\!-\!1]\mathbf{F}_1[n\!-\!1]\right)= \mathrm{Span}\left(\mathbf{H}_{32}[n\!-\!1]\mathbf{F}_2[n\!-\!1]\right).\label{const3}
\end{eqnarray}
There exists $\frac{M}{2}$-dimensional subspace orthogonal to $\frac{M}{2}$-dimensional space spanned by the aligned inter-relay interference signals. Therefore, the $\mathrm{rank}$-$\frac{M}{2}$ amplifying matrix, $\mathbf{F}_3[n]$, can be developed on the orthogonal subspace of the inter-relay interference signals. If the $\mathrm{rank}$-$\frac{M}{2}$ amplifying matrices at the relay side satisfy the conditions in (\ref{const1and2}) and (\ref{const3}), the inter-relay interference can be perfectly canceled for all time slots.


Now, we develop the linear precoder and amplifying filters to maximize the above system achievable sum-rate under the zero inter-relay interference condition. As observed in
(\ref{const1and2}) and (\ref{const3}), we note that the amplifying filters for successive time slots are affected by each other. The achievable sum-rate, $I[n,n\!-\!1]$, is also concatenated by
the achievable sum-rates of the previous and next time slot since the design criterion of $\mathbf{F}_i[n]$ is related to $\mathbf{F}_j[n\!-\!1]$ and $\mathbf{F}_j[n\!+\!1]$ for $j\neq i$.
Ideally, to compute the optimum filters maximizing the achievable sum-rate, we should solve a joint sequential optimization problem with parameter sets, $\{\mathbf{F}_i[n]|\forall n,\forall i\}$ and $\{\mathbf{T}[n]|\forall n\}$, which is given by
\begin{eqnarray}
\max_{\{\mathbf{F}_i[n]|\forall n,\forall i\},\{\mathbf{T}[n]|\forall n\}}\sum_{\forall n} I[n,n\!-\!1].\label{opt_prob}
\end{eqnarray}
However, we cannot easily compute this sequential solution since we need noncausal channel knowledge for all time slots and the channel is varying over a coherence time as well. In addition, this kind of joint optimization problem requires a huge amount of memories and makes an implementation complicated. Therefore, we now propose suboptimal filter designs which aim at finding symbol-by-symbol linear filters for the source and relays.

For convenience, when we calculate the linear filters by solving the optimization problem, we omit a time index of variables and make new definitions which depend on the time slot and, which are listed in what follows:
\begin{eqnarray}
\mathbf{T}[n]\!\!&\!\!=\!\!&\!\!\begin{cases}
\mathbf{T}_e, & \textrm{even }n\\
\mathbf{T}_o, & \textrm{odd }n,\\
\end{cases}
\;\;\;\;p_t[n] = \begin{cases} p_e, & \textrm{even }n\\ p_o, & \textrm{odd }n\\ \end{cases} \nonumber\\
\mathbf{F}_i[n]\!\!&\!\!=\!\!&\!\!\mathbf{F}_i, \;\;\;\;p_i[n]=p_i, \;\; i=1,2,3, \nonumber\\
\boldsymbol\Sigma_i[n]= \boldsymbol\Sigma_i \!\!&\!\!=\!\!&\!\!\begin{cases} p_e\mathbf{H}_{i\!S}\mathbf{T}_e\mathbf{T}_e^{\mathsf{H}}\mathbf{H}_{i\!S}^{\mathsf{H}}+\sigma_i^2\mathbf{I}_{\!M}, & \textrm{even }n\; (i=1,2) \\
p_o\mathbf{H}_{i\!S}\mathbf{T}_o\mathbf{T}_o^{\mathsf{H}}\mathbf{H}_{i\!S}^{\mathsf{H}}+\sigma_i^2\mathbf{I}_{\!M}, & \textrm{odd }n\; (i=3) \\ \end{cases} \nonumber\\
\tilde{\mathbf{H}}[n,n\!-\!1]\!\!&\!\!=\!\!&\!\!
\begin{cases}
\mathbf{H}_o=\sqrt{p_3}\mathbf{H}_{\!D3}\mathbf{F}_3\mathbf{H}_{3\!S}\mathbf{T}_o, & \textrm{even }n \\
\mathbf{H}_e=\sum_{i=1}^2\sqrt{p_i}\mathbf{H}_{\!Di}\mathbf{F}_i\mathbf{H}_{i\!S}\mathbf{T}_e, & \textrm{odd }n\\
\end{cases}\nonumber\\
\boldsymbol\Sigma_{\tilde{\mathbf{z}}}[n]\!\!&\!\!=\!\!&\!\!\begin{cases}
\boldsymbol\Sigma_o = \sigma_3^2p_3\mathbf{H}_{\!D3}\mathbf{F}_3\mathbf{F}_3^{\mathsf{H}}\mathbf{H}_{\!D3}^{\mathsf{H}}+\sigma_{\!D}^2\mathbf{I}_{\!M}, &\textrm{even }n \\
\boldsymbol\Sigma_e = \sum_{i=1}^2 \sigma_i^2p_i\mathbf{H}_{\!Di}\mathbf{F}_i\mathbf{F}_i^{\mathsf{H}}\mathbf{H}_{\!Di}^{\mathsf{H}}+\sigma_{\!D}^2\mathbf{I}_{\!M},
&\textrm{odd }n\\ \end{cases} \nonumber
\end{eqnarray}
\begin{eqnarray}
\mathbf{E}[n,n-1]\!\!&\!\!=\!\!&\!\!\begin{cases} \mathbf{E}_o=
\left(\mathbf{I}_{\!\frac{M}{2}}+p_o\mathbf{H}_o^{\mathsf{H}}\boldsymbol\Sigma_o^{\mathsf{-1}}\mathbf{H}_o\right)^{\mathsf{-1}}, & \textrm{even }n \\
\mathbf{E}_e= \left(\mathbf{I}_{\!M}+p_e\mathbf{H}_e^{\mathsf{H}}\boldsymbol\Sigma_e^{\mathsf{-1}}\mathbf{H}_e\right)^{\mathsf{-1}}, & \textrm{odd }n. \\ \end{cases} \nonumber\\
I[n,n-1]\!\!&\!\!=\!\!&\!\!\begin{cases} I_o = \log_2\det\mathbf{E}_o^{-1}, & \textrm{even }n \\
I_e = \log_2\det\mathbf{E}_e^{-1}, & \textrm{odd }n \\ \end{cases} \nonumber
\end{eqnarray}
We obviously note that all variables with the subscript $e$ are related to the transmission over $S$-$(R_1,R_2)$-$D$ link during two time slots from an even to an odd time slot and the rest with the subscript $o$ are related to the transmission over $S$-$R_3$-$D$ link during two time slots from an odd to an even time slot.

\subsection{Iterative Source/Relay Filter Design For Slow Fading}
\label{sec_design_slow}

Let us first consider the filter design for slow fading channel when the channel gain is random but remains constant, $\mathbf{H}_{ij}[n]=\mathbf{H}_{ij}$ for all $n$.
At this time, the design of the linear filters to
maximize the achievable sum-rate in (\ref{opt_prob}) is equivalent to jointly optimize the linear filters for two time slots since the optimizations for every
two time slots are the same regardless of a time index. Therefore, the linear filters for the source and relay nodes for every even time slot and every odd time
slot remain constant and the design of linear filters for both time slots is related to each other regardless of a time index.
In this case, the rate-maximization problem in~(\ref{opt_prob}) can be reformulated as a joint optimization problem with parameters for even time slot and odd time slot, which is given by
\begin{eqnarray}
&\displaystyle\max_{\mathbf{F}_1,\mathbf{F}_2,\mathbf{F}_3,\mathbf{T}_e,\mathbf{T}_o} & \frac{1}{2}\left(\log_2\det\mathbf{E}_o^{\mathsf{-1}}+\log_2\det\mathbf{E}_e^{\mathsf{-1}}\right)\nonumber\\
&\textrm{s.t.}& \mathrm{Span}\!\left(\mathbf{F}_3\right)\perp\mathrm{Span}\!\left(\mathbf{H}_{31}\mathbf{F}_1\right)=\mathrm{Span}\!\left(\mathbf{H}_{32}\mathbf{F}_2\right)\label{opt2}\\
&&\mathrm{Span}\!\left(\mathbf{H}_{13}\mathbf{F}_3\right)\perp\mathrm{Span}\!\left(\mathbf{F}_1\right), \;\mathrm{Span}\!\left(\mathbf{H}_{23}\mathbf{F}_3\right)\perp\mathrm{Span}\!\left(\mathbf{F}_2\right).\nonumber
\end{eqnarray}
To solve this problem, each node requires global channel state information (CSI) which can be acquired at the beginning of transmission.
This is not a convex optimization which is difficult to be solved by standard optimization tools. In order to find a suboptimal solution, we define an amplifying matrix for each relay as a product of two $\mathrm{rank}$-$\frac{M}{2}$ matrices, which is given by
\begin{eqnarray}
\mathbf{F}_i=\mathbf{B}_i\mathbf{W}_i^\mathsf{H},\label{f_define}
\end{eqnarray}
where $\mathbf{B}_i\in\mathbb{C}^{M\times\frac{M}{2}}$ and $\mathbf{W}_i\in\mathbb{C}^{M\times\frac{M}{2}}$ are respectively referred to as a forward matrix and a backward matrix in this paper.
Then, the IA constraint in~(\ref{const3}) to cancel the inter-relay interference at $R_3$ can be rewritten as
\begin{eqnarray}
\mathrm{Span}(\mathbf{W}_3)\perp \mathrm{Span}(\mathbf{H}_{31}\mathbf{B}_1)=\mathrm{Span}(\mathbf{H}_{32}\mathbf{B}_2).\label{rel_w3_b1b2}
\end{eqnarray}
We make the following structure:
$\mathbf{H}_{31}\mathbf{B}_1=\mathbf{U}_{b}\boldsymbol\phi_1$ and $\mathbf{H}_{32}\mathbf{B}_2=\mathbf{U}_{b}\boldsymbol\phi_2$, where $\mathbf{U}_b\in\mathbb{C}^{M\times\frac{M}{2}}$ is a basis matrix which spans the aligned interference subspace and $\boldsymbol\phi_i\in\mathbb{C}^{\frac{M}{2}\times\frac{M}{2}}$ is an arbitrary matrix. Now we can rewrite the backward and forward matrices as
\begin{eqnarray}
\mathbf{W}_3=\mathbf{U}_b^\bot\boldsymbol\psi_3, \textrm{ and }\mathbf{B}_i=\mathbf{H}_{3i}^{-1}\mathbf{U}_{b}\boldsymbol\phi_i,\;\;i=1,2,
\label{relation_b}
\end{eqnarray}
where we define $\mathbf{Z}^{\bot}=\mathbf{I}_M-\mathbf{Z}(\mathbf{Z}^\mathsf{H}\mathbf{Z})^{-1}\mathbf{Z}^\mathsf{H}$ for an arbitrary $\mathbf{Z}$ and $\boldsymbol\psi_3\in\mathbb{C}^{M\times\frac{M}{2}}$ is an arbitrary matrix. Since $\mathbf{W}_3$ should be orthogonal to the aligned interference signals, it is projected onto the orthogonal subspace of $\mathbf{U}_b$.

On the other hand, in order to cancel the inter-relay interference at $R_1$ and $R_2$, the following conditions should be met:
\begin{eqnarray}
\mathrm{Span}\!\left(\mathbf{H}_{i3}\mathbf{B}_3\right)\perp \mathrm{Span}\!\left(\mathbf{W}_i\right),\;\;i=1,2.
\label{rel_w1w2_b3}
\end{eqnarray}
Both conditions in (\ref{rel_w1w2_b3}) are equivalently represented as $\mathrm{Span}\!\left(\mathbf{B}_3\right)\perp \mathrm{Span}\!\left(\mathbf{H}_{13}^{\mathsf{H}}\mathbf{W}_1\right)$ and $\mathrm{Span}\!\left(\mathbf{B}_3\right)\perp \mathrm{Span}\!\left(\mathbf{H}_{23}^{\mathsf{H}}\mathbf{W}_2\right)$. Since both $\mathbf{H}_{13}^{\mathsf{H}}\mathbf{W}_1$ and $\mathbf{H}_{23}^{\mathsf{H}}\mathbf{W}_2$ are orthogonal to $\mathbf{B}_3$ and span an $\frac{M}{2}$-dimensional subspace, two matrices should be the matrices lying on the same subspace which is presented as $\mathrm{Span}\!\left(\mathbf{H}_{13}^{\mathsf{H}}\mathbf{W}_1\right)=\mathrm{Span}\!\left(\mathbf{H}_{23}^{\mathsf{H}}\mathbf{W}_2\right)$. Similarly to the structures of the forward matrices, $\mathbf{B}_1$ and $\mathbf{B}_2$, $\mathbf{W}_1$ and $\mathbf{W}_2$ have the structure as $\mathbf{H}_{13}^{\mathsf{H}}\mathbf{W}_1=\mathbf{U}_w\boldsymbol\psi_1$ and $\mathbf{H}_{23}^{\mathsf{H}}\mathbf{W}_2=\mathbf{U}_w\boldsymbol\psi_2$, where $\mathbf{U}_w\in\mathbb{C}^{M\times\frac{M}{2}}$ is a basis matrix and $\boldsymbol\psi_i\in\mathbb{C}^{\frac{M}{2}\times\frac{M}{2}}$ is an arbitrary matrix. Using the notations, we can rewrite the forward and backward matrices as
\begin{eqnarray}
\mathbf{B}_3=\mathbf{U}_{w}^{\bot}\boldsymbol\phi_3, \textrm{ and }\mathbf{W}_i=\mathbf{H}_{i3}^{\mathsf{-H}}\mathbf{U}_{w}\boldsymbol\psi_i,\;\;i=1,2,
\label{relation_w}
\end{eqnarray}
where $\boldsymbol\phi_3\in\mathbb{C}^{M\times\frac{M}{2}}$ is an arbitrary matrix. The amplifying matrix filters for the relays are represented as
\begin{eqnarray}
\begin{array}{rcl}
\mathbf{F}_i=\mathbf{H}_{3i}^{-1}\mathbf{U}_{b}\mathbf{G}_i\mathbf{U}_{w}^H\mathbf{H}_{i3}^{-1},\;\;i=1,2, \;\;\textrm{and}\;\;
\mathbf{F}_3=\mathbf{U}_w^{\bot}\mathbf{G}_3\mathbf{U}_b^{\bot},
\end{array}\label{amp_filter}
\end{eqnarray}
where $\mathbf{G}_i=\boldsymbol\phi_i\boldsymbol\psi_i^\mathsf{H}$ is an arbitrary matrix. Using new definitions in (\ref{amp_filter}), (\ref{opt2}) can be reformulated as
\begin{eqnarray}
\max_{\mathbf{U}_b,\mathbf{U}_w,\mathbf{G}_1,\mathbf{G}_2,\mathbf{G}_3,\mathbf{T}_e,\mathbf{T}_o}\;  \frac{1}{2}\left(\log_2\det\mathbf{E}_o^{\mathsf{-1}}+\log_2\det\mathbf{E}_e^{\mathsf{-1}}\right).\nonumber
\end{eqnarray}

In order to maximize the above achievable sum-rate, we consider an iterative algorithm using the subgradient method which is a first-order optimization to always guarantee finding a local minimum of an objective function\footnote{It is very simple and easy to use though exhibiting very slow convergence in the worst case. Briefly reviewing the operation of this method, the derivative of the objective function is given by $\partial f(\mathbf{Z},\mathbf{Z}^\mathsf{*})/\partial \mathbf{Z}^{\mathsf{*}}$, where $f(\mathbf{Z},\mathbf{Z}^{\mathsf{*}})$ is an objective function with respect to $\mathbf{Z}$ and $\mathbf{Z}^\mathsf{*}$. The $k$th iteration of the method can be formulated as $\mathbf{Z}^{[k+1]}=\mathbf{Z}^{[k]}+\mu^{[k]}(\partial f(\mathbf{Z}^{[k]},\mathbf{Z}^{[k]*})/\partial \mathbf{Z}^\mathsf{*})$, where $\mu^{[k]}$ is a step size parameter. In this paper, the step size parameter should be determined by Armijo's rule~\cite{armijo} guaranteeing $f(\mathbf{Z}^{[k+1]})\leq f(\mathbf{Z}^{[k]})$. We determine $\mu^{[k]}=\nu^m$ where $m$ is the smallest integer such that $f(\mathbf{Z}^{[k]}+\nu^m\partial f(\mathbf{Z}^{[k]},\mathbf{Z}^{[k]*})/\partial \mathbf{Z}^\mathsf{*})\leq f(\mathbf{Z}^{[k]})+\zeta\nu^m\| \partial f(\mathbf{Z}^{[k]},\mathbf{Z}^{[k]*})/\partial \mathbf{Z}^\mathsf{*}\|_F^2$ for $\zeta, \nu \in (0,1)$. We set $\zeta=0.2$ and $\nu=0.5$ for numerical results in the paper.}. Therefore, we should first find the partial derivatives of the objective function with respect to $\mathbf{U}_b^{\mathsf{*}}$, $\mathbf{U}_w^{\mathsf{*}}$, $\mathbf{G}_1^{\mathsf{*}}$, $\mathbf{G}_2^{\mathsf{*}}$, $\mathbf{G}_3^{\mathsf{*}}$, $\mathbf{T}_e^{\mathsf{*}}$ and $\mathbf{T}_o^{\mathsf{*}}$, respectively to compute the direction in which it increases the fastest for each iteration. When we define the objective function as $f_1=\frac{1}{2}(I_o+I_e)$, the partial derivatives of $f_1$ with respect to the matrices at the relay side can be computed as\footnote{In Appendix~\ref{app1}, we describe the derivation of finding the partial derivatives in detail.}
\begin{eqnarray}
\frac{\partial f_1}{\partial \mathbf{U}_b^\mathsf{*}}\!&\!=\!&\!\frac{p_e}{2\ln2}\sum_{i=1}^2\mathbf{H}_{3i}^{-\mathsf{H}}\boldsymbol\Psi_i
\mathbf{H}_{i3}^{-\mathsf{H}}
\mathbf{U}_w\mathbf{G}_i^\mathsf{H}-\frac{p_o}{2\ln2}\mathbf{U}_b^{\bot}\left(\mathbf{G}_3^\mathsf{H}\mathbf{U}_w^{\bot}\boldsymbol\Psi_3+
\boldsymbol\Psi_3^{\mathsf{H}}\mathbf{U}_w^{\bot}\mathbf{G}_3\right)\mathbf{U}_b^{\dag},\label{parUb}\\
\frac{\partial f_1}{\partial \mathbf{U}_w^\mathsf{*}}\!&\!=\!&\!\frac{p_e}{2\ln2}\sum_{i=1}^2\mathbf{H}_{i3}^{-1}\boldsymbol\Psi_i^{\mathsf{H}}
\mathbf{H}_{3i}^{-1}\mathbf{U}_b\mathbf{G}_i-\frac{p_o}{2\ln2}\mathbf{U}_w^{\bot}\left(\boldsymbol\Psi_3\mathbf{U}_b^{\bot}
\mathbf{G}_3^\mathsf{H}+\mathbf{G}_3\mathbf{U}_b^{\bot}\boldsymbol\Psi_3^{\mathsf{H}}\right)\mathbf{U}_w^{\dag},\label{parUw}\\
\frac{\partial f_1}{\partial \mathbf{G}_i^*}\!&\!=\!&\!\frac{p_e}{2\ln2}\mathbf{U}_b^\mathsf{H}\mathbf{H}_{3i}^{-\mathsf{H}}\boldsymbol\Psi_i
\mathbf{H}_{i3}^{-\mathsf{H}}\mathbf{U}_w,\;\; i=1,2,\label{parG1andG2}\\
\frac{\partial f_1}{\partial
\mathbf{G}_3^*}\!&\!=\!&\!\frac{p_o}{2\ln2}\mathbf{U}_w^{\bot}\boldsymbol\Psi_3\mathbf{U}_b^{\bot},\label{parG3}
\end{eqnarray}
where we define $\mathbf{Z}^\dag=\mathbf{Z}(\mathbf{Z}^\mathsf{H}\mathbf{Z})^{-1}$ for arbitrary $\mathbf{Z}$, and
\begin{eqnarray}
\boldsymbol\Omega_i\!&\!=\!&\!\mathbf{H}_{\!Di}^\mathsf{H}\boldsymbol\Sigma_e^{-1}\mathbf{H}_e\mathbf{E}_e\left(\mathbf{T}_e^\mathsf{H}
\mathbf{H}_{iS}^\mathsf{H}-\sigma_i^2\sqrt{p_i}\mathbf{H}_e^\mathsf{H}\boldsymbol\Sigma_e^{-1}\mathbf{H}_{\!Di}\mathbf{F}_i\right), \;\;i=1, 2, \nonumber\\
\boldsymbol\Omega_3\!&\!=\!&\!\mathbf{H}_{D3}^\mathsf{H}\boldsymbol\Sigma_o^{-1}\mathbf{H}_o\mathbf{E}_o\left(\mathbf{T}_o^\mathsf{H}\mathbf{H}_{3S}^\mathsf{H}
-\sigma_3^2\sqrt{p_3}\mathbf{H}_o^\mathsf{H}\boldsymbol\Sigma_o^{-1}\mathbf{H}_{D3}\mathbf{F}_3\right),\nonumber\\
\boldsymbol\Psi_i\!&\!=\!&\!\sqrt{p_i}\left[\boldsymbol\Omega_i-\frac{p_i}{P_R}Re\left(\mathrm{tr}
\left\{\mathbf{F}_i^\mathsf{H}\boldsymbol\Omega_i\right\}\right)\mathbf{F}_i\boldsymbol\Sigma_i\right], \;\; i=1, 2, 3.\nonumber
\end{eqnarray}

From now, we should find the partial derivatives of $f_1$ with respect to $\mathbf{T}_e^\mathsf{*}$ and $\mathbf{T}_o^\mathsf{*}$ at $S$. In~\cite[Eq. 19]{leekj10}, the partial derivative of achievable sum-rate with respect to a precoding matrix at the source has been obtained for one and two-way relaying systems using multiple MIMO relays. We utilize this result to find the partial derivatives for both $\mathbf{T}_e^\mathsf{*}$ and $\mathbf{T}_o^\mathsf{*}$ which are given by
\begin{eqnarray}
\frac{\partial f_1}{\partial \mathbf{T}_e^*}\!\!&\!\!=\!\!&\!\!-\frac{p_e^2}{2P_S\ln2}\mathrm{tr}\left\{\mathbf{H}_e^\mathsf{H}\boldsymbol\Sigma_e\mathbf{H}_e
\mathbf{E}_e\right\}\mathbf{T}_e+\frac{p_e}{2\ln2}\sum_{i=1}^2\sqrt{p_i}\mathbf{H}_{i\!S}^\mathsf{H}\mathbf{F}_i^\mathsf{H}
\mathbf{H}_{\!Di}^\mathsf{H}
\boldsymbol\Sigma_e\mathbf{H}_e\mathbf{E}_e\nonumber\\
\!\!&\!\!-\!\!&\!\!\sum_{i=1}^2\frac{p_e^2p_i\sqrt{p_i}}{2P_R\ln2}Re\left(\mathrm{tr}\left\{\mathbf{F}_i^\mathsf{H}\boldsymbol\Omega_i
\right\}\right)\left(\boldsymbol\Phi_i-\frac{p_e}{P_S}\mathrm{tr}\left\{\mathbf{T}_e^\mathsf{H}\boldsymbol\Phi_i\mathbf{T}_e
\right\}\mathbf{I}_M\right)\mathbf{T}_e,\label{parTe}\\
\frac{\partial f_1}{\partial \mathbf{T}_o^*}\!\!&\!\!=\!\!&\!\!-\frac{p_o^2}{2P_S\ln2}\mathrm{tr}\left\{\mathbf{H}_o^\mathsf{H}
\boldsymbol\Sigma_o\mathbf{H}_o\mathbf{E}_o\right\}\mathbf{T}_o+\frac{p_o\sqrt{p_3}}{2\ln2}\mathbf{H}_{3\!S}^\mathsf{H}
\mathbf{F}_3^\mathsf{H}\mathbf{H}_{\!D3}^\mathsf{H}\boldsymbol\Sigma_o\mathbf{H}_o\mathbf{E}_o\nonumber\\
\!\!&\!\!-\!\!&\!\!\frac{p_o^2p_3\sqrt{p_3}}{2P_R\ln2}Re\left(\mathrm{tr}\left\{\mathbf{F}_3^\mathsf{H}\boldsymbol\Omega_3
\right\}\right)\left(\boldsymbol\Phi_3-\frac{p_o}{P_S}\mathrm{tr}\{\mathbf{T}_o^\mathsf{H}\boldsymbol\Phi_3\mathbf{T}_o\}
\mathbf{I}_M\right)\mathbf{T}_o,\label{parTo}
\end{eqnarray}
where we define $\boldsymbol\Phi_i=\mathbf{H}_{i\!S}^\mathsf{H}\mathbf{F}_i^\mathsf{H}\mathbf{F}_i\mathbf{H}_{i\!S}$ for $i=1$, 2, 3. Using the partial derivatives with respect to the related matrices at the source and relay side, we propose an iterative algorithm applying the subgradient method for each matrix sequentially. The basic idea of the method is to take a step along the direction of the gradient with respect to each matrix for each iteration and repeat the iteration until approaching a local maximum of the achievable sum-rate. We describe the mode of operation for the proposed iterative algorithm in the following. In this algorithm, $\mu_i$ for $i\in\{b,w,1,2,3,e,o\}$ is a step size parameter allowed to change at every
iteration and $\epsilon$ is a precision factor to terminate the iterative procedure.
\vspace{0.1in}
{\hrule height 1pt}
\vspace{0.05in}
{\small
{\bf \normalsize Iterative Algorithm I \label{algo1}}
\vspace{0.02in}
\hrule
\vspace{0.08in}
\hspace{-0.2in} Initialization
\begin{enumerate}
\item[1)] Initialize the matrices, $\mathbf{U}_b^{[k]}$, $\mathbf{U}_w^{[k]}$, $\mathbf{G}_1^{[k]}$, $\mathbf{G}_2^{[k]}$, $\mathbf{G}_3^{[k]}$, $\mathbf{T}_e^{[k]}$ and $\mathbf{T}_o^{[k]}$ for $k=0$.
\end{enumerate}
Iteration
\begin{enumerate}
\item[2)] Compute the partial derivative, ${\partial f_1(\mathbf{U}_b^{[k]})}/{\partial \mathbf{U}_b^\mathsf{*}}$, and update the matrix, $\mathbf{U}_b^{[k+1]}= \mathbf{U}_b^{[k]} + \mu_b^{[k]} \partial f_1(\mathbf{U}_b^{[k]})/\partial \mathbf{U}_b^*$.
\item[3)] Compute the partial derivative, ${\partial f_1(\mathbf{U}_w^{[k]})}/{\partial \mathbf{U}_w^\mathsf{*}}$, and update the matrix, $\mathbf{U}_w^{[k+1]}= \mathbf{U}_w^{[k]} + \mu_w^{[k]} {\partial f_1(\mathbf{U}_w^{[k]})}/{\partial \mathbf{U}_w^\mathsf{*}}$.
\item[4)] Compute the partial derivative, ${\partial f_1(\mathbf{G}_i^{[k]})}/{\partial \mathbf{G}_i^\mathsf{*}}$, and update each matrix, $\mathbf{G}_i^{[k+1]}= \mathbf{G}_i^{[k]} + \mu_i^{[k]} {\partial f_1(\mathbf{G}_i^{[k]})}/{\partial \mathbf{G}_i^\mathsf{*}}$ for $i=1$, 2, 3.
\item[5)] Compute the partial derivative, ${\partial f_1(\mathbf{T}_i^{[k]})}/{\partial \mathbf{T}_i^\mathsf{*}}$, and update each matrix, $\mathbf{T}_i^{[k+1]}= \mathbf{T}_i^{[k]} + \mu_i^{[k]} {\partial f_1(\mathbf{T}_i^{[k]})}/{\partial \mathbf{T}_i^\mathsf{*}}$ for $i\in\{e, o\}$.
\item[6)] If $f_1^{[k+1]}-f_1^{[k]}\leq \epsilon$, stop iteration. Otherwise, $k\leftarrow k+1$ and repeat 2)-5).
\end{enumerate}
Results
\begin{enumerate}
\item[7)] Output the matrices, $\mathbf{T}_i^{[k\!+\!1]}$ for $i\in\{e,o\}$ and $\mathbf{F}_i^{[k\!+\!1]}$ for $i=1$, 2, 3.
\end{enumerate}}
{\hrule height 1pt}

\subsection{Alternately Iterative Source/Relay Filter Design For Fast Fading}
\label{sec_design_fast}

\subsubsection{Scenario 1: Flat Fading Per Two Time Slots}
\label{sec_design_fast2}

Now we consider the filter design for the block fading channel which is often assumed in cooperative systems or relaying systems. It is usually assumed in relaying systems that the channel remains constant during two hops, which the transmission stages from source to relay and from relay to destination are called first phase and second phase, respectively. Likewise, we assume that the channel matrices during two consecutive time slots over $S$-($R_1,R_2$)-$D$ link remain constant, which is presented as $\mathbf{H}_{ij}[n\!-\!1]=\mathbf{H}_{ij}[n]$ for $i,j\in\{S,D,1,2,3\}$ and $i\neq j$ for odd $n$.
In this fading scenario, we develop a distributed alternate relaying system without exchanging channel information or using feedback information to cancel the inter-relay interference. In order not to utilize feedback channels to report the forward channel information to $S$, we only consider the design of the amplifying filters at the relay side. Since the transmit precoding filter at $S$ is not dependent on the channel characteristics, we simply set $\mathbf{T}[n]=\mathbf{I}_M$ for even $n$ and $\mathbf{T}[n]=\mathbf{I}_{M,1:\frac{M}{2}}$ for odd $n$.

We note that each relay only knows its local channel information, that is, $R_i$ has only backward, forward and inter-relay channel information, $\mathbf{H}_{i\!S}[n]$, $\mathbf{H}_{\!Di}[n]$, and $\mathbf{H}_{ij}[n]$ for $j\neq i$. Each relay can estimate backward/forward channel information by receiving training signals broadcasted by $S$ and $D$, respectively and inter-relay channel information by eavesdropping pilot signals sent to $D$ by $R_3$. Although each relay knows local CSI, it is necessary to cancel the inter-relay interference so that the relays forward the desired message to $D$. Without loss of generality, let us consider odd $n$. From (\ref{y_relay}) and (\ref{f_define}), for the $n\!-\!1$th and $n$th time slots, the received signals at $R_i$ can be written as
\begin{eqnarray}
\mathbf{y}_i[n\!-\!1]\!\!&\!\!=\!\!&\!\!\mathbf{H}_{i\!S}[n\!-\!1]\mathbf{T}[n\!-\!1]\mathbf{s}[n\!-\!1]+\mathbf{H}_{i3}[n\!-\!1]
\mathbf{B}_3[n\!-\!1]\mathbf{W}_3^\mathsf{H}[n\!-\!1]\mathbf{y}_3[n\!-\!2]+\mathbf{z}_i[n\!-\!1],\;\;i=1,2,\nonumber\\
\mathbf{y}_3[n]\!\!&\!\!=\!\!&\!\!\mathbf{H}_{3\!S}[n]\mathbf{T}[n]\mathbf{s}[n]+\sum_{i=1}^2
\mathbf{H}_{3i}[n]\mathbf{B}_i[n]\mathbf{W}_i^\mathsf{H}[n]\mathbf{y}_i[n\!-\!1]+\mathbf{z}_3[n].\nonumber
\end{eqnarray}
In order to cancel the interference at $R_1$ and $R_2$ at the $n$th time slot, (\ref{rel_w1w2_b3}) should be satisfied, that is, $\mathbf{W}_i^\mathsf{H}[n]\mathbf{H}_{i3}[n\!-\!1]\mathbf{B}_{3}[n\!-\!1]=\mathbf{0}$ for $i=1$, 2.
Meanwhile, $R_3$ should remove the interference signals so that $R_3$ can forward its $\frac{M}{2}$ desired messages, that is,
$\mathbf{W}_3^\mathsf{H}[n\!+\!1]\mathbf{H}_{3i}[n]\mathbf{B}_{i}[n]=\mathbf{0}$ for $i=1$, 2.
Due to the reciprocity of channels between relays and constant block fading during the $n\!-\!1$th to $n$th time slots, it is shown that $\mathbf{H}_{3i}[n]=\mathbf{H}_{i3}^\mathsf{T}[n-1]$ for $i=1$, 2. Using this equality, the condition for interference cancelation at $R_3$ can be rewritten as
$\mathbf{W}_3^\mathsf{H}[n\!+\!1]\mathbf{H}_{i3}^\mathsf{T}[n\!-\!1]\mathbf{B}_{i}[n]=
\mathbf{0}$ for $i=1$, 2.
Both conditions for interference cancelation at all relays can be met at once by setting
\begin{eqnarray}
\begin{array}{rcl}
\mathbf{W}_i[n]\!&\!=\!&\!\bar{\mathbf{U}}_i\\
\mathbf{B}_i[n]\!&\!=\!&\!\mathbf{W}_i^*[n]\boldsymbol\xi_i[n] \\ \mathbf{W}_3[n\!+\!1]\!&\!=\!&\!\mathbf{B}_3^*[n\!-\!1]\boldsymbol\xi_3[n\!+\!1],
\end{array}\label{relation_nofd}
\end{eqnarray}
where $\bar{\mathbf{U}}_i$ is an orthonormal matrix which spans the space orthogonal to $\mathbf{H}_{i3}[n\!-\!1]\mathbf{B}_3[n\!-\!1]$ and  $\boldsymbol\xi_i[\cdot]\in\mathbb{C}^{\frac{M}{2}\times\frac{M}{2}}$ is a matrix determined by local optimization at $R_i$.
If $R_3$ decides the forward matrix, $\mathbf{B}_3[n\!-\!1]$, for its desired signal, $R_1$ and $R_2$ can find the backward matrices, $\mathbf{W}_1[n]$ and $\mathbf{W}_2[n]$, orthogonal to $\mathbf{H}_{13}[n\!-\!1]\mathbf{B}_3[n\!-\!1]$ and $\mathbf{H}_{23}[n\!-\!1]\mathbf{B}_3[n\!-\!1]$, respectively. The forward matrices, $\mathbf{B}_1[n]$ and $\mathbf{B}_2[n]$, can be also obtained by using $\mathbf{W}_1[n]$ and $\mathbf{W}_2[n]$, while $R_3$ exploits $\mathbf{B}_3[n\!-\!1]$ to determine the backward matrix, $\mathbf{W}_3[n\!+\!1]$, orthogonal to $\mathbf{H}_{31}[n]\mathbf{B}_1[n]$ and $\mathbf{H}_{32}[n]\mathbf{B}_2[n]$. With this setting, each relay does not have to know other relays' channel information to align interference signals but its own local channel information. Now we should first optimize the forward matrix, $\mathbf{B}_3[n\!-\!1]$, for the $n\!-\!1$th time slot to maximize the achievable sum-rate over $S$-$R_3$-$D$ link. Based on it, $\boldsymbol\xi_1[n]$ and $\boldsymbol\xi_2[n]$ are locally optimized after determining backward filters. First of all, we note that the backward matrix for the $n\!-\!1$th time slot, $\mathbf{W}_3[n\!-\!1]$, is already given since it is determined by previous forward matrix. $R_3$ computes the amplifying matrix, $\mathbf{F}_3[n\!-\!1]$, to maximize the achievable rate based on the backward/forward channel information, $\mathbf{H}_{3S}[n\!-\!2]$ and $\mathbf{H}_{D3}[n\!-\!1]$, for the $n\!-\!1$th time slot as well as the backward channel information, $\mathbf{H}_{3S}[n\!+\!1]$, for the $n\!+\!1$th time slot since $\mathbf{B}_3[n\!-\!1]$ is related to $\mathbf{W}_3[n+1]$ as in (\ref{relation_nofd}).
The MSE matrix for data vector at $D$ over $S$-$R_3$-$D$ link in (\ref{basic_E}) for the $n\!-\!1$th time slot can be rewritten as
\begin{eqnarray}
\mathbf{E}[n\!-\!1,n\!-\!2]=\left(\mathbf{I}[n\!-\!1]+\sqrt{p_t[n\!-\!2]}
\tilde{\mathbf{H}}^\mathsf{H}[n\!-\!1,n\!-\!2]\boldsymbol\Sigma_{\tilde{z}}^{-1}[n\!-\!1]
\tilde{\mathbf{H}}[n\!-\!1,n\!-\!2]\right)^{-1},\nonumber
\end{eqnarray}
where $\tilde{\mathbf{H}}[n\!-\!1,n\!-\!2]=\sqrt{p_3[n\!-\!1]}\mathbf{H}_{\!D3}[n\!-\!1]
\mathbf{B}_3[n\!-\!1]\mathbf{W}_3^\mathsf{H}[n\!-\!1]\mathbf{H}_{3\!S}[n\!-\!2]
\mathbf{T}[n\!-\!2]$. We emphasize that $\mathbf{B}_3[n\!-\!1]$ also affects the performance of the data vector for the $n\!+\!1$th time slot\footnote{$\mathbf{B}_3[n\!-\!1]$ is related to $\mathbf{W}_1[n]$ and $\mathbf{W}_2[n]$ as in (\ref{relation_nofd}) but we cannot consider joint optimization of them since we assume that $R_3$ only knows its local CSI.}. At the $n\!-\!1$th time slot, $R_3$ cannot estimate the forward channel, $\mathbf{H}_{\!D3}[n\!+\!1]$ but can estimate the backward channel, $\mathbf{H}_{3\!S}[n]$. We thus introduce a new objective function to measure the performance over $S$-$R_3$ link when we use the transmit precoding matrix, $\mathbf{T}[n]$, and the backward matrix, $\mathbf{W}_3[n+1]$.
Multiplying the backward matrix by the received signal at $R_3$ for the odd time slot given in (\ref{y_relay}), the postprocessing signal can be
calculated as
\begin{eqnarray}
\mathbf{y}_p=\mathbf{W}_3^\mathsf{H}[n\!+\!1]\mathbf{y}_3[n]=\sqrt{p_t[n]}
\mathbf{H}_p\mathbf{s}[n]+\mathbf{z}_p,\label{y_p}
\end{eqnarray}
where we define $\mathbf{H}_p=\mathbf{W}_3^\mathsf{H}[n\!+\!1]\mathbf{H}_{3\!S}[n]\mathbf{T}[n]$ and $\mathbf{z}_p=\mathbf{W}_3^\mathsf{H}[n\!+\!1]\mathbf{z}_3[n]$ with $\boldsymbol\Sigma_p=\mathbb{E}[\mathbf{z}_p\mathbf{z}_p^\mathsf{H}]=\sigma_3^2
\mathbf{W}_3^\mathsf{H}[n\!+\!1]\mathbf{W}_3[n\!+\!1]$. Adapting MMSE filter for the postprocessing received signal, its MSE matrix and achievable sum-rate can be evaluated as
$\mathbf{E}_p=\left(\mathbf{I}[n]+p_t[n]\mathbf{H}_p^\mathsf{H}\boldsymbol\Sigma_p^{-1}
\mathbf{H}_p\right)^{-1}$
and $I_p=\log_2\det\mathbf{E}_p^{-1}$, where $\mathbf{H}_p=\mathbf{W}_3^\mathsf{H}[n\!+\!1]\mathbf{H}_{3\!S}[n]\mathbf{T}[n]=
\boldsymbol\xi_3^\mathsf{H}[n\!+\!1]\mathbf{B}_3^\mathsf{T}[n\!-\!1]
\mathbf{H}_{3\!S}[n]\mathbf{T}[n]$. We arbitrarily determine $\boldsymbol\xi_3[n\!+\!1]=
\mathbf{I}_{\!\frac{M}{2}}$ since $\boldsymbol\xi_3[n\!+\!1]$ is irrelative to $\mathbf{E}_p$. From now, we omit the time index for convenience and use the simple notations listed in the previous section and, using $\mathbf{B}_3$, the MSE matrix for the $n$th time slot\footnote{We note that we put $(\cdot)^\prime$ on the channel matrix for the $n$th time slot to distinguish it from that for the $n\!-\!2$th time slot.} can be represented as $\mathbf{E}_p=\left(\mathbf{I}_{\!\frac{M}{2}}+\frac{p_o}{\sigma_3^2}\mathbf{T}_o^\mathsf{H}
{\mathbf{H}}_{3\!S}^{\prime\mathsf{H}}\mathbf{B}_3^*\mathbf{B}_3^{\dag\mathsf{T}}
{\mathbf{H}}_{3\!S}^\prime\mathbf{T}_o\right)^{-1}$.
We formulate the problem to maximize the achievable sum-rate as
${\displaystyle \max_{\mathbf{B}_3}}\;  \frac{1}{2}(I_o+I_p)$.
Now in order to utilize an iterative algorithm based on the subgradient method, we should compute the partial derivative of the achievable sum-rate, $f_4=\frac{1}{2}(I_o+I_p)$ with respect to $\mathbf{B}_3^*$ which is given by
\begin{eqnarray}
\frac{\partial f_4}{\partial \mathbf{B}_3^*}=\frac{p_o}{2\ln 2}\left(\boldsymbol\Psi_3\mathbf{W}_3+\mathbf{B}_3^\bot
\boldsymbol\Upsilon_3^{\mathsf{T}}\mathbf{B}_3^\dag\right),\label{parB3_nofd}
\end{eqnarray}
where we here note that $\boldsymbol \Upsilon_3$ is not a function of $\mathbf{H}_{3S}$ but $\mathbf{H}_{3S}^\prime$.
Given the backward matrix, $\mathbf{W}_3$, $R_3$ computes the forward matrix, $\mathbf{B}_3$, by using the method of steepest ascent as shown in the distributed algorithm at the end of this section. For the $n$th time slot, $R_1$ and $R_2$ compute the amplifying matrix orthogonal to the inter-relay interference signal from $R_3$ based on the relation in (\ref{relation_nofd}), which is given by $\mathbf{F}_i=\bar{\mathbf{U}}_i^*\boldsymbol\xi_i\bar{\mathbf{U}}_i^\mathsf{H}$ for $i=1$, 2. Since $\bar{\mathbf{U}}_i$ can computed by the received signal from the previous time slot, each relay should find $\boldsymbol\xi_i$ to maximize the achievable sum-rate. In order to find the amplifying matrices for the relays which have only local channel information, MMSE and zeroforcing-based filter design has been proposed in~\cite{oyman06}. However, these filters cannot be applied in this scenario because it is not successful to cancel the interference among $M$ data streams via $\frac{M}{2}$ received signals with the postprocessing backward channel matrix, $\bar{\mathbf{U}}_i^\mathsf{H}\mathbf{H}_{i\!S}$. Since we assume in this scenario that the CSI exchange between relays is not possible, we try to find each $\boldsymbol\xi_i$ to maximize the individual mutual information via $R_1$ and $R_2$, respectively. We assume that the received signal is given by $\mathbf{y}_{\!Di}=\sqrt{p_ep_i}\mathbf{H}_{\!Di}\mathbf{F}_i(\mathbf{H}_{i\!S}\mathbf{T}_e\mathbf{s}+
\mathbf{z}_i)+\mathbf{z}_D$. We compute the amplifying matrix to maximize the mutual information of single relay channel, $f_{ei} = \log_2\det\mathbf{E}_{ei}^{-1}$, where $\mathbf{E}_{ei}=\left(\mathbf{I}_{\!M}+p_ep_i\mathbf{H}_{i\!S}^\mathsf{H}\mathbf{F}_i^\mathsf{H}\mathbf{H}_{\!Di}^\mathsf{H}
\boldsymbol\Sigma_{ei}^{-1}\mathbf{H}_{\!Di}\mathbf{F}_i\mathbf{H}_{i\!S}\right)^{-1}$ and $\boldsymbol\Sigma_{ei}=\sigma_i^2 p_i\mathbf{H}_{\!Di}\mathbf{F}_i\mathbf{F}_i^\mathsf{H}
\mathbf{H}_{\!Di}^\mathsf{H}+\sigma_D^2\mathbf{I}_{\!M}$. We consider an iterative algorithm using the method of steepest ascent\footnote{There have been several algorithms to optimize the relay filter for single MIMO relaying channel~\cite{medina07,rong09}. We can also apply these methods to find the amplifying filters.} and then the partial derivatives of $f_{ei}$ with respect to $\boldsymbol\xi_i^*$ can be computed as
\begin{eqnarray}
\frac{\partial f_{ei}}{\partial \boldsymbol\xi_i^*}=\frac{p_e}{2\ln 2}\bar{\mathbf{U}}_i^\mathsf{T}\boldsymbol\Psi_{ei}\bar{\mathbf{U}}_i.\label{parxi12}
\end{eqnarray}
In (\ref{parxi12}), we define $\boldsymbol\Psi_{ei}=\sqrt{p_i}\left(\boldsymbol\Omega_{ei}-\frac{p_i}{P_R}Re\left(\mathrm{tr}\{\mathbf{F}_i^\mathsf{H}\boldsymbol\Omega_{ei}\}\right)
\mathbf{F}_i\boldsymbol\Sigma_i\right)$, where $\mathbf{H}_{ei}=\sqrt{p_i}\mathbf{H}_{\!Di}\mathbf{F}_i\mathbf{H}_{i\!S}$ and $\boldsymbol\Omega_{ei}=\mathbf{H}_{\!Di}^H\boldsymbol\Sigma_{ei}^{-1}\mathbf{H}_{ei}\mathbf{E}_{ei}\left(\mathbf{H}_{i\!S}^\mathsf{H}
-\sigma_i^2\sqrt{p_i}\mathbf{H}_{ei}^\mathsf{H}\boldsymbol\Sigma_{ei}^{-1}\mathbf{H}_{\!Di}\mathbf{F}_i\right)$.
$R_i$ for $i=1$, 2 can compute the amplifying matrix, $\mathbf{F}_i=\bar{\mathbf{U}}_i^*\boldsymbol\xi_i\bar{\mathbf{U}}_i^\mathsf{H}$ for the $n$th time slot as shown in the following algorithm. Finally, we find the distributed iterative algorithm alternately utilized for each time slot based on (\ref{relation_nofd}).

\vspace{0.1in}
{\hrule height 1pt}
\vspace{0.05in}
{\small 
{\bf \normalsize Distributed Algorithm \label{algo3}}
\vspace{0.02in}
\hrule
\vspace{0.08in}

\hspace{-0.15in}{\bf Case I}: even $n$,
\begin{itemize}
\item[]Initialization
\begin{enumerate}
\item[1)] Given the matrices, $\mathbf{T}_o=\mathbf{I}_{M,1:\frac{M}{2}}$ and $\mathbf{W}_3$, initialize the matrices, $\mathbf{B}_3^{[k]}$ for $k=0$.
\end{enumerate}
\item[]Iteration
\begin{enumerate}
\item[2)] Compute the partial derivative, ${\partial f_4(\mathbf{B}_3^{[k]})}/{\partial \mathbf{B}_3^\mathsf{*}}$, and update the matrix, $\mathbf{B}_3^{[k+1]}= \mathbf{B}_3^{[k]} + \mu_w^{[k]} {\partial f_4(\mathbf{B}_3^{[k]})}/{\partial \mathbf{B}_3^\mathsf{*}}$.
\item[3)] If $f_4^{[k+1]}-f_4^{[k]}\leq \epsilon$, stop iteration. Otherwise, $k\leftarrow k+1$ and repeat 2).
\end{enumerate}
\item[]Results
\begin{enumerate}
\item[4)] Output the amplifying matrix, $\mathbf{F}_3^{[k\!+\!1]}$, for the $n$th time slot and the backward matrix, $\mathbf{W}_3=\mathbf{B}_3^{[k\!+\!1]*}$, for the $n\!+\!2$th time slot.
\end{enumerate}
\end{itemize}
{\bf Case II}: odd $n$,
\begin{itemize}
\item[]Initialization for the $i$th relay ($i=1$,2)
\begin{enumerate}
\item[1)] Find the matrix, $\bar{\mathbf{U}}_i$, orthogonal to $\mathbf{H}_{i3}\mathbf{B}_3$.
\item[2)] Initialize the matrix, $\boldsymbol\xi_i^{[k]}$
\end{enumerate}
\item[]Iteration
\begin{enumerate}
\item[3)] Compute the partial derivative, ${\partial f_{ei}(\boldsymbol\xi_i^{[k]})}/{\partial \boldsymbol\xi_i^\mathsf{*}}$, and update the matrix, $\boldsymbol\xi_i^{[k+1]}= \boldsymbol\xi_i^{[k]} + \mu_i^{[k]} \partial f_{ei}(\boldsymbol\xi_i^{[k]})/\partial \boldsymbol\xi_i^*$.
\item[4)] If $f_{ei}^{[k+1]}-f_{ei}^{[k]}\leq \epsilon$, stop iteration. Otherwise, $k\leftarrow k+1$ and repeat 3).
\end{enumerate}
\item[]Results
\begin{enumerate}
\item[5)] Output the matrices, $\mathbf{F}_i^{[k\!+\!1]}=\bar{\mathbf{U}}_i^*\boldsymbol\xi_i^{[k\!+\!1]}\bar{\mathbf{U}}_i^\mathsf{H}$ for the transmission during odd time slot.
\end{enumerate}
\end{itemize}}

{\hrule height 1pt}
\vspace{0.1in}


\subsubsection{Scenario 2: Flat Fading Per One Time Slot}
\label{sec_design_fast1}

In this section, we first consider a
filter design for a block fading scenario in which the channel is varying every time slot, i.e., $\mathbf{H}_{ij}[n]\neq\mathbf{H}_{ij}[n\!-\!1]$ for any $n$. For any $n$th data symbols, the joint optimization of the transmit precoding filter, $\mathbf{T}[n]$, at $S$ and amplifying filter, $\mathbf{F}_i[n\!+\!1]$, at the relays cannot be applied simultaneously since the forward channel for the next time slot cannot be estimated at present. Recalling (\ref{f_define}), the backward filter, $\mathbf{W}_3[n]$, is jointly optimized with the filters, $\mathbf{B}_1[n\!-\!1]$ and $\mathbf{B}_2[n\!-\!1]$, for the $n\!-\!1$th messages since it cancels the inter-relay interference signal induced by the $n\!-\!1$th message in (\ref{rel_w3_b1b2}). On the other hand, $\mathbf{B}_3[n]$ is jointly optimized with $\mathbf{W}_1[n\!+\!1]$ and $\mathbf{W}_2[n\!+\!1]$ due to the constraints in (\ref{rel_w1w2_b3}). At this time, it is required to know global CSI to perform the joint optimization.

First, we focus on the optimization for even $n$ to find $\mathbf{T}[n]$, $\mathbf{W}_1[n\!+\!1]$, $\mathbf{W}_2[n\!+\!1]$ and $\mathbf{B}_3[n]$. We note that the transmit precoding filter, $\mathbf{T}[n\!-\!1]$, and the backward filter, $\mathbf{W}_3[n]$, are given through the optimization for the previous odd time slot. We need to introduce a new objective function since we cannot estimate the channel matrix, $\mathbf{H}_{\!D1}[n\!+\!1]$ and
$\mathbf{H}_{\!D2}[n\!+\!1]$, for the next odd time slot due to the nature of block fading and then cannot use the MSE matrix, $\mathbf{E}[n\!+\!1, n]$. The received signals at $R_1$ and $R_2$ for the even time slot are given in (\ref{y_relay}) and multiplying the backward matrices by them yields
\begin{eqnarray}
\mathbf{y}_c=\left[\!\!
               \begin{array}{c}
                 \mathbf{W}_1^\mathsf{H}[n\!+\!1]\mathbf{y}_1[n] \\
                 \mathbf{W}_2^\mathsf{H}[n\!+\!1]\mathbf{y}_2[n] \\
               \end{array}
            \!\! \right]=\sqrt{p_t[n]}\mathbf{H}_c\mathbf{s}[n]+\mathbf{z}_c,\nonumber
\end{eqnarray}
where the compound channel matrix and noise are defined as
\begin{eqnarray}
\mathbf{H}_c=\left[\!\!
               \begin{array}{c}
                 \mathbf{W}_1^\mathsf{H}[n\!+\!1]\mathbf{H}_{1\!S}[n] \\
                 \mathbf{W}_2^\mathsf{H}[n\!+\!1]\mathbf{H}_{2\!S}[n] \\
               \end{array}
            \!\! \right]\mathbf{T}[n],\;\;\;
\mathbf{z}_c=\left[\!\!
               \begin{array}{c}
                 \mathbf{W}_1^\mathsf{H}[n\!+\!1]\mathbf{z}_1[n] \\
                 \mathbf{W}_2^\mathsf{H}[n\!+\!1]\mathbf{z}_2[n] \\
               \end{array}
            \!\! \right].\nonumber
\end{eqnarray}
We define the covariance matrix of noise vector $\mathbf{z}_c$ as
\begin{eqnarray}
\boldsymbol\Sigma_c=\mathbb{E}[\mathbf{z}_c\mathbf{z}_c^\mathsf{H}]=
\left[\!\!
  \begin{array}{cc}
    \sigma_1^2\mathbf{W}_1^\mathsf{H}[n\!+\!1]\mathbf{W}_1[n\!+\!1]\!\! &\!\! \mathbf{0} \\
    \mathbf{0}\!\! &\!\! \sigma_2^2\mathbf{W}_2^\mathsf{H}[n\!+\!1]\mathbf{W}_2[n\!+\!1] \\
  \end{array}
\!\!\right].\nonumber
\end{eqnarray}
We use the MSE matrix of this compound received signal as an objective function in part for the optimization of linear filters for even time slot. We assume MMSE linear filter for this compound signal, $\mathbf{W}_c = (p_t[n]\mathbf{H}_c\mathbf{H}_c^\mathsf{H}+\boldsymbol\Sigma_c)^{-1}\sqrt{p_t[n]}\mathbf{H}_c$,
 although it is not practically used in the system. The MSE matrix for this compound signal can be computed as $\mathbf{E}_c=(\mathbf{I}[n]+p_t[n]\mathbf{H}_c^\mathsf{H}\boldsymbol\Sigma_c^{-1}
\mathbf{H}_c)^{-1}.$
From now, we use the simple notations listed in the previous section for convenience. We define the achievable sum-rate for the compound signal as $I_c=\log_2\det\mathbf{E}_c^{-1}$, where
\begin{eqnarray}
\mathbf{E}_c=\left(\mathbf{I}_M+\sum_{i=1}^2\frac{p_e}{\sigma_i^2}\mathbf{T}_e^\mathsf{H}
\mathbf{H}_{i\!S}^\mathsf{H}\mathbf{W}_i^\dag\mathbf{W}_i^\mathsf{H}\mathbf{H}_{i\!S}
\mathbf{T}_e\right)^{-1}.\nonumber
\end{eqnarray}
We formulate the problem to maximize the total achievable sum-rate for even time slot as
\begin{eqnarray}
\begin{array}{cl}
 \displaystyle \max_{\mathbf{W}_1,\mathbf{W}_2,\mathbf{B}_3,\mathbf{T}_e} & \frac{1}{2}(I_o+I_c) \\
  \mathrm{s.t.} & \mathrm{Span}(\mathbf{H}_{i3}\mathbf{B}_3)\perp\mathrm{Span}(\mathbf{W}_i), \;i=1,2,
\end{array}\nonumber
\end{eqnarray}
where $\mathbf{T}_o$ and $\mathbf{W}_3$ are given from the previous slot. Recalling (\ref{relation_w}), this problem is reformulated as
\begin{eqnarray}
\begin{array}{cc}
  \displaystyle \max_{\mathbf{U}_w,\boldsymbol\phi_3,\mathbf{T}_e} & \frac{1}{2}\left(
  \log_2\det\mathbf{E}_o^{-1}+\log_2\det\mathbf{E}_c^{-1}\right),
\end{array}\nonumber
\end{eqnarray}
where $\boldsymbol\psi_1$ and $\boldsymbol\psi_2$ are irrelevant to the optimization since they cannot affect the MSE matrix of the compound signal and therefore we simply set $\boldsymbol\psi_1 = \boldsymbol\psi_2 = \mathbf{I}_{\!\frac{M}{2}}$. In the same manner as the previous proposed algorithm, we utilize the iterative algorithm based on the subgradient method. We compute the partial derivative of $f_2$ with respect to each matrix taking a step to a local maximum for each iteration, where we define $f_2=\frac{1}{2}(I_o+I_c)$. The partial derivatives of
$f_o$ with respect to $\mathbf{U}_w^*$, $\boldsymbol\phi_3^*$ and $\mathbf{T}_e^*$ are given by\footnote{In Appendix~\ref{app2}, we describe the derivation of finding the partial derivatives in detail.}
\begin{eqnarray}
\frac{\partial f_2}{\partial \mathbf{U}_w^*}\!&\!=\!&\!\frac{p_e}
{2\ln 2}\sum_{i=1}^2\frac{1}{\sigma_i^2}\mathbf{H}_{i3}^{-1}\mathbf{W}_i^\bot
\boldsymbol\Upsilon_i\mathbf{W}_i^\dag-\frac{p_o}{2\ln 2}\mathbf{U}_w^\bot\left(
\boldsymbol\Psi_3\mathbf{W}_3\boldsymbol\phi_3^\mathsf{H}+\boldsymbol\phi_3
\mathbf{W}_3^\mathsf{H}\boldsymbol\Psi_3^{\mathsf{H}}\right)\mathbf{U}_w^\dag,
\label{parUw2}\end{eqnarray}
\begin{eqnarray}
\frac{\partial f_2}{\partial \boldsymbol\phi_3^*}\!&\!=\!&\!\frac{p_o}{2\ln 2}\mathbf{U}_w^\bot
\boldsymbol\Psi_3\mathbf{W}_3,\label{par_phi3_2}\\
\frac{\partial f_2}{\partial \mathbf{T}_e^*}\!&\!=\!&\!\frac{p_e}{2\ln 2}\left(\sum_{i=1}^2
-\frac{p_e}{\sigma_i^2P_S}\mathrm{tr}\left\{\boldsymbol\Upsilon_i
\mathbf{W}_i^\dag\mathbf{W}_i^\mathsf{H}\right\}\mathbf{T}_e+\frac{1}{\sigma_i^2}
\mathbf{H}_{i\!S}^\mathsf{H}\mathbf{W}_i^\dag\mathbf{W}_i^\mathsf{H}\mathbf{H}_{i\!S}
\mathbf{T}_e\mathbf{E}_c\right),\label{parTe2}
\end{eqnarray}
where we denote $\boldsymbol\Upsilon_i=\mathbf{H}_{i\!S}\mathbf{T}_e\mathbf{E}_c
\mathbf{T}_e^\mathsf{H}\mathbf{H}_{i\!S}^\mathsf{H}$ for $i=1$, 2. Finally, we describe the proposed iterative algorithm for even time slot based on successively applying the subgradient method for each optimizing matrix in {\bf Case I} of the following algorithm at the end of section.

On the other hand, the optimization for odd $n$ is to find $\mathbf{T}[n]$, $\mathbf{B}_1[n]$, $\mathbf{B}_2[n]$ and $\mathbf{W}_3[n\!+\!1]$ maximizing the achievable sum-rate for odd time slot when the backward filters and transmit precoding filter, $\mathbf{W}_1[n]$, $\mathbf{W}_2[n]$ and $\mathbf{T}[n\!-\!1]$, are given from the previous time slot. In the same way as the optimization for even time slot, we cannot use the MSE matrix for
$S$-$R_3$-$D$ link since we cannot know the channel between $R_3$ and $D$ at present. Recalling the postprocessing recieved signal in (\ref{y_p}), adapting MMSE filter for the postprocessing received signal, its MSE matrix and achievable sum-rate can be evaluated as
$\mathbf{E}_p=\left(\mathbf{I}[n]+p_t[n]\mathbf{H}_p^\mathsf{H}\boldsymbol\Sigma_p^{-1}
\mathbf{H}_p\right)^{-1}$
and $I_p=\log_2\det\mathbf{E}_p^{-1}$. Using simple notations without time indices, we can formulate the optimization problem for odd time slot such as
\begin{eqnarray}
\begin{array}{cl}
  \displaystyle\max_{\mathbf{B}_1,\mathbf{B}_2,\mathbf{W}_3,\mathbf{T}_o} & \frac{1}{2}
  (I_e+I_p) \\
  \mathrm{s.t.} & \mathrm{Span}(\mathbf{W}_3)\perp\mathrm{Span}(\mathbf{H}_{31}\mathbf{B}_1)=
  \mathrm{Span}(\mathbf{H}_{32}\mathbf{B}_2),
\end{array}\nonumber
\end{eqnarray}
where $\mathbf{T}_e$, $\mathbf{W}_1$ and $\mathbf{W}_2$ are given from the previous time slot. Recalling (\ref{relation_b}), the above problem can be reformulated as
\begin{eqnarray}
\begin{array}{cc}
  \displaystyle\max_{\mathbf{U}_b,\boldsymbol\phi_1,\boldsymbol\phi_2,\mathbf{T}_o} &
  \frac{1}{2}(\log_2\det\mathbf{E}_e^{-1}+\log_2\det \mathbf{E}_p^{-1}),
\end{array}\nonumber
\end{eqnarray}
where $\mathbf{E}_p=\left(\mathbf{I}_{\!\frac{M}{2}}+\frac{p_o}{\sigma_3^2}
\mathbf{T}_o^\mathsf{H}\mathbf{H}_{3\!S}^\mathbf{H}\mathbf{W}_3^\dag
\mathbf{W}_3^\mathsf{H}\mathbf{H}_{3\!S}\mathbf{T}_o\right)^{-1}$ and we set $\boldsymbol\psi_3=\mathbf{I}_{\!M,1:\frac{M}{2}}$ since $I_p$ does not depend on $\boldsymbol\psi_3$. The partial derivatives of the achievable rate, $f_3 = \frac{1}{2} (I_e +I_p)$, with respect to $\mathbf{U}_b^*$, $\boldsymbol\phi_1^*$, $\boldsymbol\phi_2^*$ and $\mathbf{T}_o^*$ can be obtained as
\begin{eqnarray}
\frac{\partial f_3}{\partial \mathbf{U}_b^*}\!&\!=\!&\!\frac{p_e}{2\ln 2}\sum_{i=1}^2\mathbf{H}_{3i}^{-\mathsf{H}}\boldsymbol\Psi_i\mathbf{W}_i \boldsymbol\phi_i^\mathsf{H}-\frac{p_o}{2\sigma_3^2\ln 2}\mathbf{U}_b^\bot\left(\mathbf{W}_3^\bot\boldsymbol\Upsilon_3\mathbf{W}_3^\dag \boldsymbol\psi_3^\mathsf{H}+\boldsymbol\psi_3\mathbf{W}_3^{\dag\mathsf{H}} \boldsymbol\Upsilon_3\mathbf{W}_3^\bot\right)\mathbf{U}_b^\dag,
\label{parUb2}\end{eqnarray}
\begin{eqnarray}
\frac{\partial f_3}{\partial \boldsymbol\phi_i^*}\!&\!=\!&\!\frac{p_e}{2\ln 2}\mathbf{U}_b^\mathsf{H}\mathbf{H}_{3i}^{-\mathsf{H}}\boldsymbol\Psi_i
\mathbf{W}_i,\;\;i=1,2,\label{parGi2}\\
\frac{\partial f_3}{\partial \mathbf{T}_e^*}\!&\!=\!&\!\frac{p_o}{2\ln 2}\left(-\frac{p_o} {\sigma_3^2 P_S}\mathrm{tr}\left\{\boldsymbol\Upsilon_3 \mathbf{W}_3^\dag \mathbf{W}_3^\mathsf{H}\right\}\mathbf{T}_o+\frac{1}{\sigma_3^2}
\mathbf{H}_{3\!S}^\mathsf{H}\mathbf{W}_3^\dag\mathbf{W}_3^\mathsf{H}\mathbf{H}_{3\!S} \mathbf{T}_o\mathbf{E}_p\right),\label{parTo2}
\end{eqnarray}
where we denote $\boldsymbol\Upsilon_3=\mathbf{H}_{3\!S}\mathbf{T}_o\mathbf{E}_p \mathbf{T}_o^\mathsf{H}\mathbf{H}_{3\!S}^\mathsf{H}$. For odd time slot, we present the iterative algorithm in {\bf Case II} of the following algorithm to find the above matrices by using the subgradient method when $\mathbf{T}_e$, $\mathbf{W}_1$ and $\mathbf{W}_2$ are given from the previous even slot. Finally, the proposed algorithm for block fading channel is to alternately utilize the iterative algorithm to find the matrices which are used at present time slot, while the
matrices optimized at the previous time slot are fixed.

\vspace{0.1in}
{\hrule height 1pt}
\vspace{0.05in}
{\small 
{\bf \normalsize Iterative Algorithm II }
\vspace{0.02in}
\hrule
\vspace{0.08in}
\hspace{-0.15in}{\bf Case I}: even $n$,
\begin{itemize}
\item[]Initialization
\begin{enumerate}
\item[1)] Given the matrices, $\mathbf{T}_o$ and $\mathbf{W}_3$, initialize the matrices, $\mathbf{U}_w^{[k]}$, $\boldsymbol\phi_3^{[k]}$, $\mathbf{T}_e^{[k]}$ for $k=0$.
\end{enumerate}
\item[]Iteration
\begin{enumerate}
\item[2)] Compute the partial derivative, ${\partial f_2(\mathbf{U}_w^{[k]})}/{\partial \mathbf{U}_w^\mathsf{*}}$, and update the matrix, $\mathbf{U}_w^{[k+1]}= \mathbf{U}_w^{[k]} + \mu_w^{[k]} {\partial f_2(\mathbf{U}_w^{[k]})}/{\partial \mathbf{U}_w^\mathsf{*}}$.
\item[3)] Compute the partial derivative, ${\partial f_2(\boldsymbol\phi_3^{[k]})}/{\partial \boldsymbol\phi_3^*}$, and update the matrix, $\boldsymbol\phi_3^{[k+1]}= \boldsymbol\phi_3^{[k]} + \mu_3^{[k]} {\partial f_2(\boldsymbol\phi_3^{[k]})}/{\partial \boldsymbol\phi_3^*}$.
\item[4)] Compute the partial derivative, ${\partial f_2(\mathbf{T}_e^{[k]})}/{\partial \mathbf{T}_e^\mathsf{*}}$, and update the matrix, $\mathbf{T}_e^{[k+1]}= \mathbf{T}_e^{[k]} + \mu_e^{[k]} {\partial f_2(\mathbf{T}_e^{[k]})}/{\partial \mathbf{T}_e^\mathsf{*}}$.
\item[5)] If $f_2^{[k+1]}-f_2^{[k]}\leq \epsilon$, stop iteration. Otherwise, $k\leftarrow k+1$ and repeat 2)-4).
\end{enumerate}
\item[]Results
\begin{enumerate}
\item[6)] Output the matrices, $\mathbf{F}_3^{[k\!+\!1]}$, $\mathbf{T}_e^{[k\!+\!1]}$, $\mathbf{W}_1^{[k\!+\!1]}$ and $\mathbf{W}_2^{[k\!+\!1]}$ for the transmission during even time slot.
\end{enumerate}
\end{itemize}
{\bf Case II}: odd $n$,
\begin{itemize}
\item[]Initialization
\begin{enumerate}
\item[1)] Given the matrices, $\mathbf{T}_e$, $\mathbf{W}_1$ and $\mathbf{W}_2$, initialize the matrices, $\mathbf{U}_b^{[k]}$, $\boldsymbol\phi_1^{[k]}$, $\boldsymbol\phi_2^{[k]}$ and $\mathbf{T}_o^{[k]}$ for $k=0$.
\end{enumerate}
\item[]Iteration
\begin{enumerate}
\item[2)] Compute the partial derivative, ${\partial f_3(\mathbf{U}_b^{[k]})}/{\partial \mathbf{U}_b^\mathsf{*}}$, and update the matrix, $\mathbf{U}_b^{[k+1]}= \mathbf{U}_b^{[k]} + \mu_b^{[k]} \partial f_3(\mathbf{U}_b^{[k]})/\partial \mathbf{U}_b^*$.
\item[3)] Compute the partial derivative, ${\partial f_3(\boldsymbol\phi_i^{[k]})}/{\partial \boldsymbol\phi_i^*}$, and update each matrix, $\boldsymbol\phi_i^{[k+1]}= \boldsymbol\phi_i^{[k]} + \mu_i^{[k]} {\partial f_3(\boldsymbol\phi_i^{[k]})}/{\partial \boldsymbol\phi_i^*}$ for $i=1$, 2.
\item[4)] Compute the partial derivative, ${\partial f_3(\mathbf{T}_o^{[k]})}/{\partial \mathbf{T}_o^\mathsf{*}}$, and update the matrix, $\mathbf{T}_o^{[k+1]}= \mathbf{T}_o^{[k]} + \mu_o^{[k]} {\partial f_3(\mathbf{T}_o^{[k]})}/{\partial \mathbf{T}_o^\mathsf{*}}$.
\item[5)] If $f_3^{[k+1]}-f_3^{[k]}\leq \epsilon$, stop iteration. Otherwise, $k\leftarrow k+1$ and repeat 2)-4).
\end{enumerate}
\item[]Results
\begin{enumerate}
\item[6)] Output the matrices, $\mathbf{F}_1^{[k\!+\!1]}$, $\mathbf{F}_2^{[k\!+\!1]}$ $\mathbf{T}_o^{[k\!+\!1]}$ and $\mathbf{W}_3^{[k\!+\!1]}$ for the transmission during odd time slot.
\end{enumerate}
\end{itemize}}

{\hrule height 1pt}
\vspace{0.2in}

\section{Numerical Results}
\label{sec_sim}

In this section, we present some selected simulation results to
compare the performance of the proposed scheme and the other schemes by Monte carlo simulations.
We consider symmetric Rayleigh fading case, that is, each element of forward and backward channel matrices is independent and identically distributed complex Gaussian random variable with zero mean and unit variance. For comparison, we here assume in our proposed scheme that $P_S=P_R=P$ and $\sigma_i^2=\sigma_D^2=\sigma^2$ for $i=1,2,3$ and define the SNR as $\rho=\frac{P}{\sigma^2}$. With our proposed protocol, we consider two different filter designs such as the iterative algorithm I in section~\ref{sec_design_slow} and the naive filter, where it simply sets $\mathbf{T}_e=\mathbf{I}_M$, $\mathbf{T}_o=\mathbf{U}_b=\mathbf{U}_w=\boldsymbol\phi_3=\boldsymbol\psi_3=\mathbf{I}_{M,1:\frac{M}{2}}$ and $\boldsymbol\phi_1=\boldsymbol\phi_2=\boldsymbol\psi_1=\boldsymbol\psi_2=\mathbf{I}_{\frac{M}{2}}$. For comparison, we consider three different schemes in conventional half-duplex mode. First, in relay cooperation scheme, all relays fully cooperate
to forward the data, i.e., they share all information and act as one relay equipped with $3M$ antennas. Secondly, we consider the best relay selection scheme which selects only one relay maximizing the sum-rate among three relays. In addition, the conventional AF relaying using a single antenna is considered. In the aforementioned three schemes, source and relay filters to maximize the sum-rate are determined by using a unified framework in~\cite{rong09}. For fair comparison of total power constraints at the source and relay during two time slots, it is assumed that $P_S=2P$ and $P_R=3P$ for the best relay selection and conventional relaying schemes as well as $P_S=2P$ and total transmit power of $3P$ over all the relays for the relay cooperation scheme unless otherwise noted. We point out that we determine each step size, $\mu_i$, for our proposed iterative schemes based on Armijo's rule~\cite{armijo} and a termination parameter as $\epsilon = 10^{-2}$, while the other iterative schemes for comparison utilize $\epsilon=10^{-4}$ for termination.

\begin{figure}[tb]
\epsfxsize 5in \centerline{\hbox{\epsffile{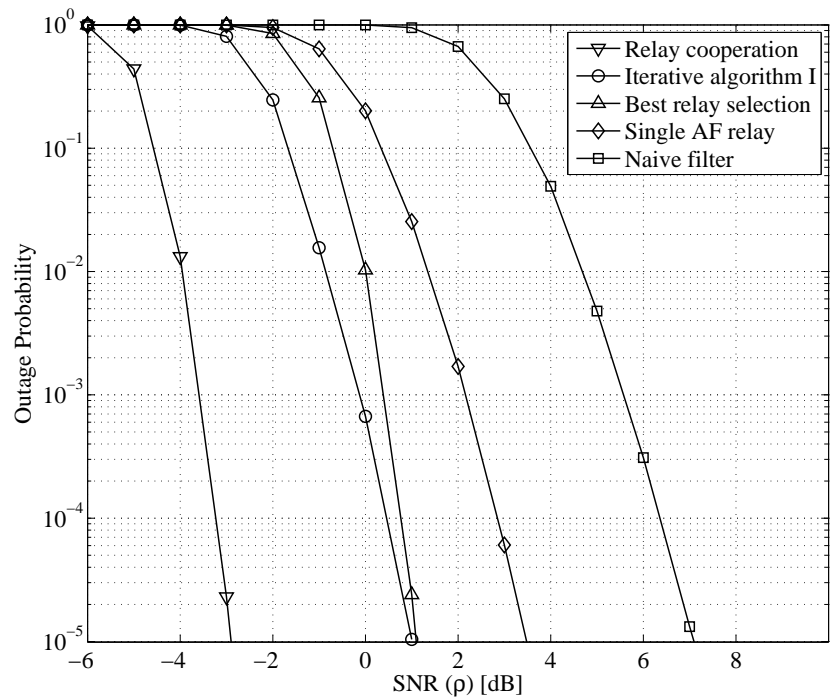}}}
\caption{Comparison of outage probability among five different schemes for slow fading channel under $M=4$ and $I_{out}=2$ [bits/s/Hz].}\label{fig2}
\end{figure}

In Fig.~\ref{fig2}, we present the outage probability of five different schemes with $M=4$ for slow fading channel. The outage probability is defined as $P_{out}=\mathrm{Pr}\left[\frac{1}{2}(I_o+I_e)\leq I_{out}\right]$ where $I_{out}$ denotes the outage threshold and we assume that $I_{out}=2$ [bits/s/Hz] in this paper. The relay cooperation scheme provides better outage probability over the other schemes since it exploits full diversity gain over $M\times 3M$ forward channel and $3M\times M$ backward channel by full relay cooperation. The best relay selection scheme has the same diversity order as the relay cooperation scheme but the different power gain since the relays are not cooperated for forwarding the data and only one relay is active during the transmission. The conventional AF relaying and naive filter give worse outage probability than our proposed iterative algorithm I for the given outage threshold, while the naive filter in the proposed protocol is even worse than the conventional scheme. However, when the source and relay filters are embedded at the nodes by using the proposed iterative algorithm I, we show from this figure that it improves the power gain significantly. Although the proposed iterative algorithm I cannot obtain the same diversity gain as relay cooperation scheme and best relay selection scheme due to the diversity-multiplexing tradeoff, it gives robust performance in terms of outage probability compared to other schemes for the SNR range of interest.

\begin{figure}[tb]
\epsfxsize 5in \centerline{\hbox{\epsffile{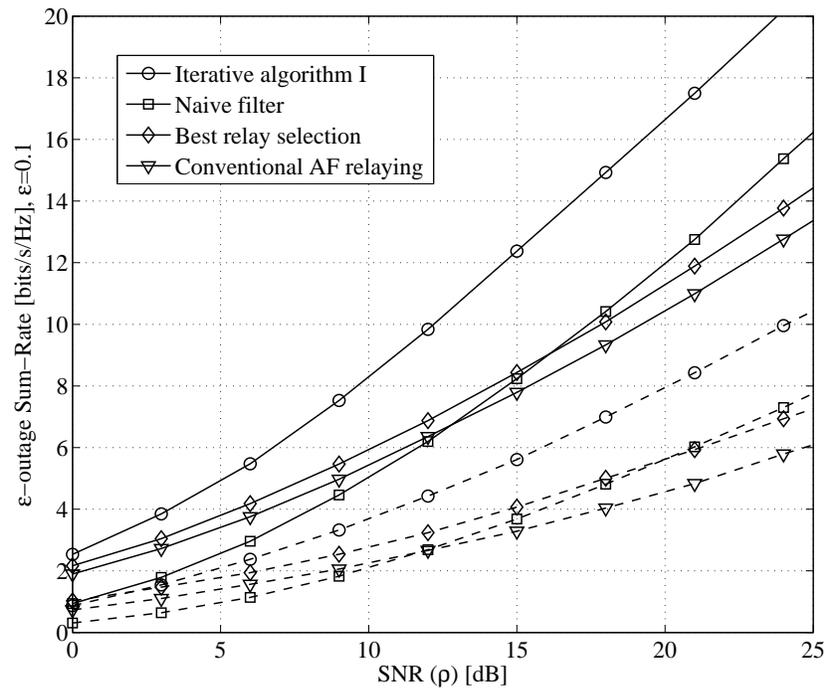}}}
\caption{Comparison of $\varepsilon$-outage sum-rate among four different schemes for slow fading channel under $M=2$ (dashed line) and 4 (solid line).}\label{fig3}
\end{figure}

Now we present $\varepsilon$-outage achievable rate of different transmission strategies with $M=2$ and $M=4$ for slow fading channel in Fig.~\ref{fig3}. The $\varepsilon$-outage achievable rate is defined as $I_{\varepsilon}=\max {I_{out}} $ subject to $P_{out}(I_{out})\leq \varepsilon$, where we set $\varepsilon=0.1$ in this paper. A naive filter provides worse outage sum-rate compared with conventional relaying schemes for low and moderate SNR region. On the other hand, our proposed scheme remains superiority over conventional schemes in the whole SNR range of interest. Hence, we know that our proposed iterative algorithm I operates suitably for slow fading environment.

\begin{figure}[t]
\centering

\subfigure[$M=2$ ]{
   \epsfxsize 5in \centerline{\hbox{\epsffile{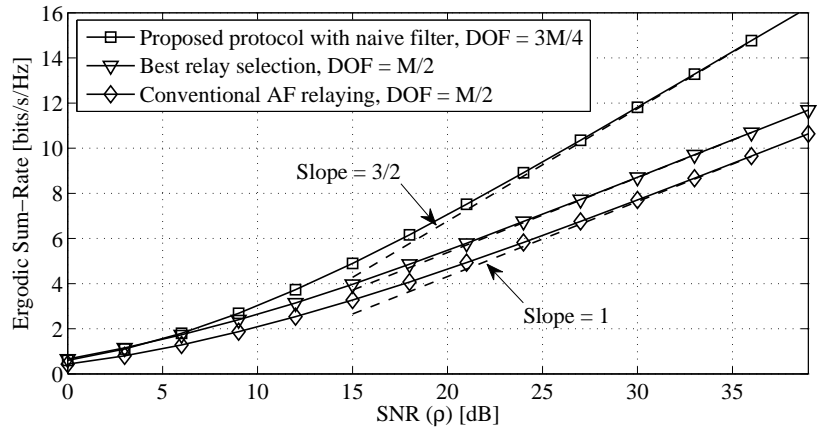}}}
   \label{slope_subfig1}
 }

 \subfigure[$M=4$]{
    \epsfxsize 5in \centerline{\hbox{\epsffile{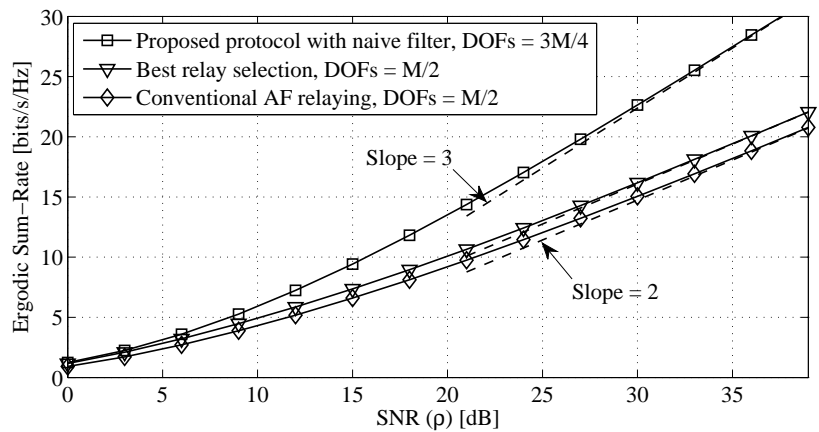}}}
   \label{slope_subfig2}
 }

\caption{Comparison of the sum-rate and capacity pre-log factor among three different schemes for block fading channel per two time slots.}\label{fig_slope}
\end{figure}

In Fig.~\ref{fig_slope}, we present the achievable DOFs and the ergodic sum-rate of three different schemes, the proposed scheme with naive filter, best relay selection scheme, and conventional AF relaying scheme for flat fading channel per two time slots. Since we verify that our proposed scheme improves the capacity pre-log factor, we simply assume that all transmission strategies simply utilize a naive filter and $P_S=P_R=P$ for power constraint at the source and relays. The capacity pre-log factor is defined as $\eta={\displaystyle\lim_{\rho\rightarrow\infty}}\frac{I(\rho)}{\log\rho}$, where $I(\rho)$ is the system sum-rate at SNR $\rho$. Given a naive filter design, we can analytically compute $\eta_e = {\displaystyle\lim_{\rho\rightarrow\infty}}\frac{I_e(\rho)}{\log\rho}=M$ for even $n$ and $\eta_o = {\displaystyle\lim_{\rho\rightarrow\infty}}\frac{I_o(\rho)}{\log\rho}=\frac{M}{2}$ for odd $n$\footnote{Although there exists a slight rate loss at the initial phase due to forwarding no data to destination, it will be of negligence to compute the DOFs for long transmission time. For instance, a rate loss during $N$ time slots is $\frac{M}{N}O(\log_2\rho)$ for initialization at odd time slot or $\frac{M}{2N}O(\log_2\rho)$ at even time slot. As $N$ increases, a rate loss goes to zero.}. The achievable DOFs in our proposed protocol is $\eta=\frac{1}{2}(\eta_e+\eta_o)=\frac{3M}{4}$. For different antenna cases, $M=2$ and $M=4$, we numerically show from these figures that our proposed scheme can obtain $\frac{3M}{4}$ DOFs, while the existing schemes using conventional half-duplex protocol achieve $\frac{M}{2}$ DOFs. Hence, we can see that our proposed scheme provides additional $\frac{M}{4}$ DOFs over conventional schemes by exploiting alternate relaying and IA.

\begin{figure}[tb]
\epsfxsize 5in \centerline{\hbox{\epsffile{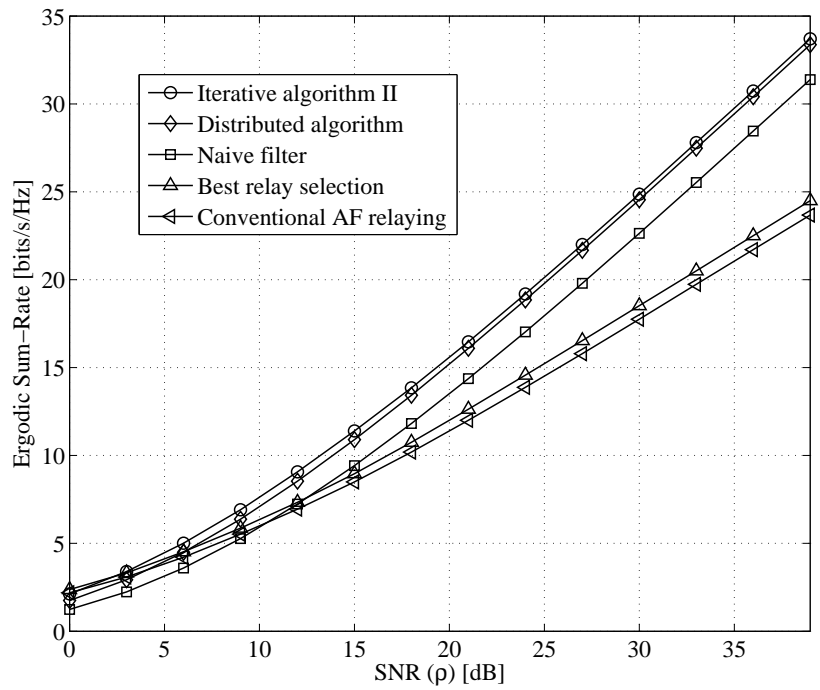}}}
\caption{Comparison of ergodic sum-rate among three different linear filters based on proposed protocol for block fading channel per two time slots.}\label{fig7}
\end{figure}

Fig.~\ref{fig7} illustrates the sum-rate performance of several linear filters for flat fading per two time slots in section \ref{sec_design_fast2}. For comparison, the transmit filters at the source are not considered also in the iterative algorithm II and we focus on the efficiency of an amplifying matrix at the relay. We can see from this figure that the proposed distributed algorithm can obtain 2 dB power gain over a naive filter. Although the distributed algorithm has a slight loss compared with the iterative algorithm II, the former can be performed locally at each relay which only requires its local CSI but the latter should require global CSI for all the relays. Therefore, we note that the proposed distributed algorithm is efficient in implementing the relays without a costly feedback load of CSI exchange between relays when the amplifying filter at the relay side is only considered.

\begin{figure}[tb]
\epsfxsize 5in \centerline{\hbox{\epsffile{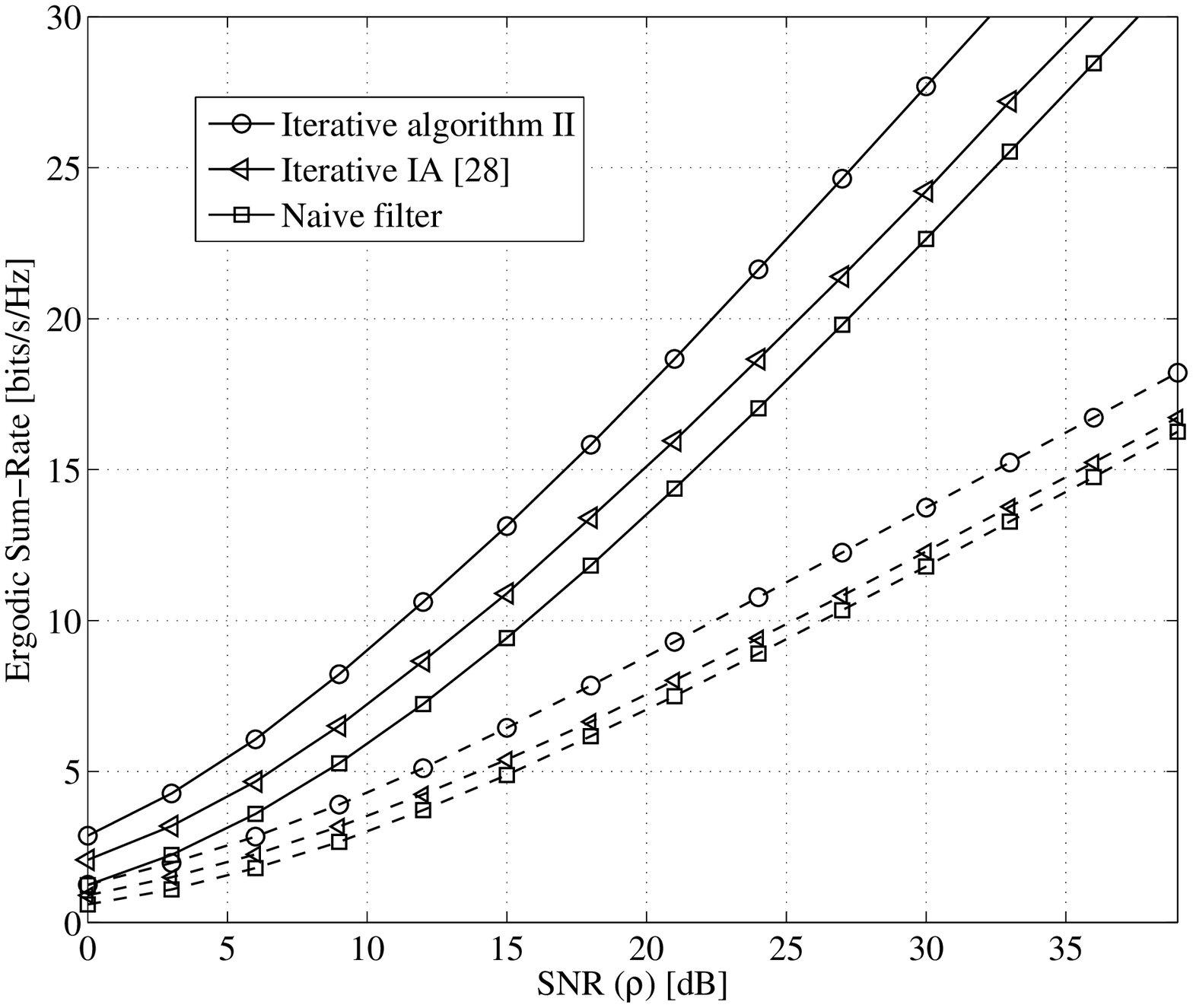}}}
\caption{Comparison of ergodic sum-rate among three different linear filters based on proposed protocol for block fading channel per one time slot in the case of $M=2$ (dashed line) and $4$ (solid line).}\label{fig5}
\end{figure}

\begin{figure}[tb]
\epsfxsize 5in \centerline{\hbox{\epsffile{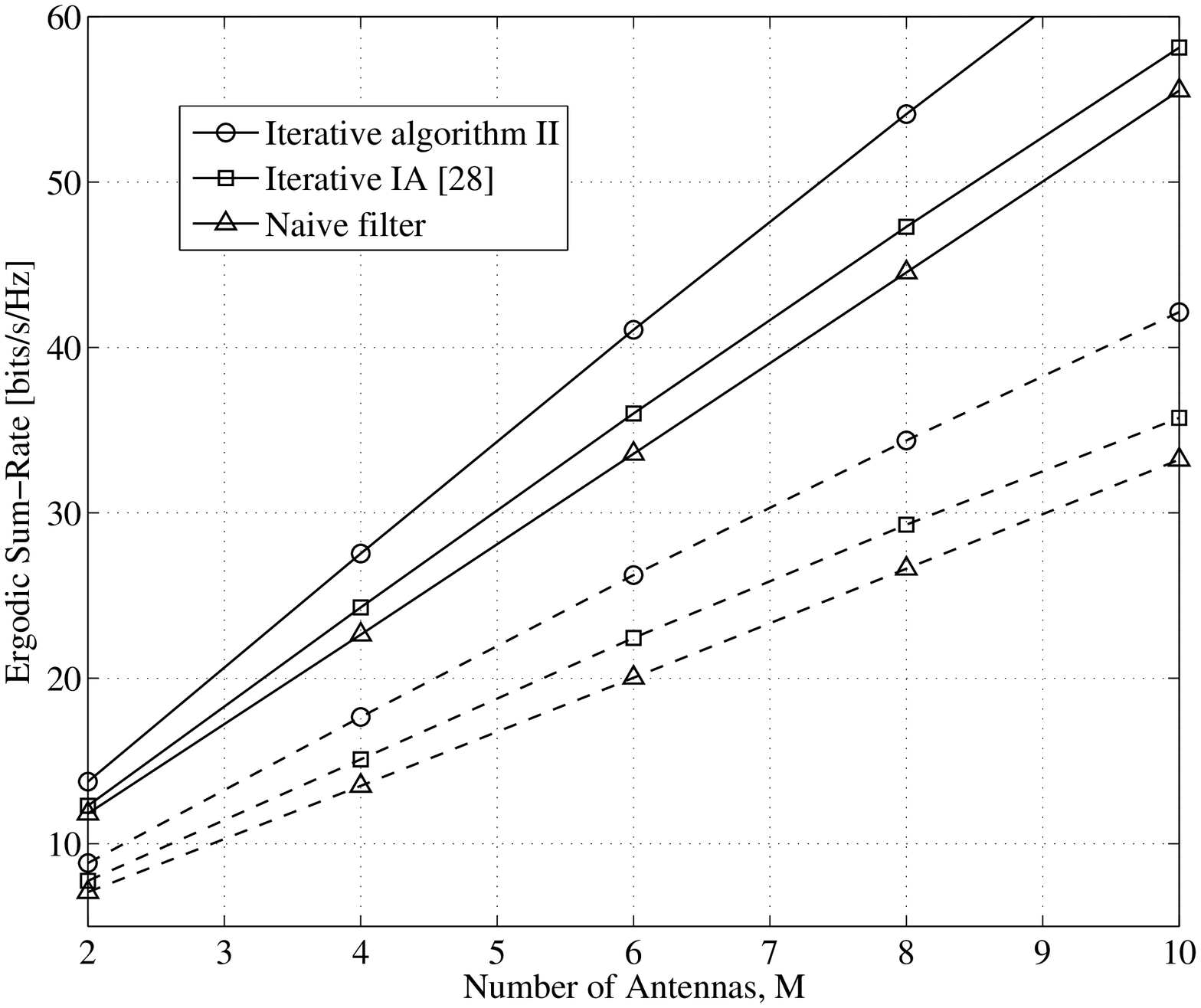}}}
\caption{Comparison of ergodic sum-rate among three different linear filters based on proposed protocol for block fading channel per one time slot in the case of $\textrm{SNR}=20$ dB (dashed line) and $\textrm{SNR}=30$ dB (solid line).}\label{fig6}
\end{figure}

In Fig.~\ref{fig5} and Fig.~\ref{fig6}, we present the sum-rate performance of three linear filters applied to our proposed protocol based on alternate relaying and IA for flat fading per one time slot in section \ref{sec_design_fast1}. For comparison, we present the performance of the proposed protocol using the previous schemes, an iterative IA~\cite{gomadam11} for inter-relay IA and an iterative algorithm~\cite{leekj10} for source and relay filter design, which is called as iterative IA in the whole figures. As shown in the case of $M=4$ in Fig.~\ref{fig5}, our proposed iterative algorithm II gives nearly 5 dB gain over naive filter and 3 dB gain over iterative IA for whole medium and high SNR region. For $M=2$, we also obtain more than 3.5 and 2.5 dB gain over those schemes, which is shown that the power gain owing to optimizing source and relay filters is increased as a function of the number of antennas, $M$. We note that our proposed scheme computes such source and relay filters that they do not only align inter-relay interference into the subspace where it maximizes the sum-rate but also optimize the sum-rate of source-to-destination channel for even and odd time slots, respectively. On the other hand, since iterative IA scheme only focuses on nulling the interference between relays regardless of the sum-rate, it cannot compute the aligned interference subspace to maximize the sum-rate and loses significant gain over our proposed scheme. A naive filter cannot even obtain any power gain resulting from maximizing source and relay filter for each time slot.

Next, we illustrate in Fig.~\ref{fig6} the sum-rate improvement of three linear filters with respect to the number of antennas, $M$, for different SNR values. As shown in this figure, the sum-rate of the proposed algorithm increases with a larger slope compared with other two schemes. We note that, as the number of antennas per node increases, our proposed scheme provides proportionally increased power gain over a naive filter. On the other hand, the power gain of iterative IA scheme over a naive filter remains nearly constant regardless of the number of antennas per node.

\begin{figure}[t]
\centering

\subfigure[Iterative algorithm I]{
   \epsfxsize 5in \centerline{\hbox{\epsffile{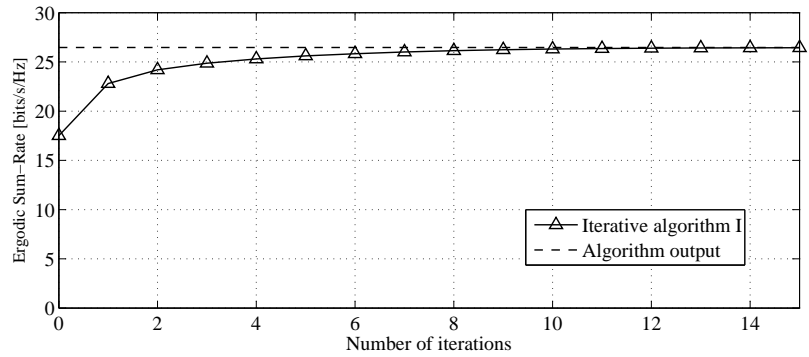}}}
   \label{fig8_1}
 }

\subfigure[Iterative algorithm II]{
    \epsfxsize 5in \centerline{\hbox{\epsffile{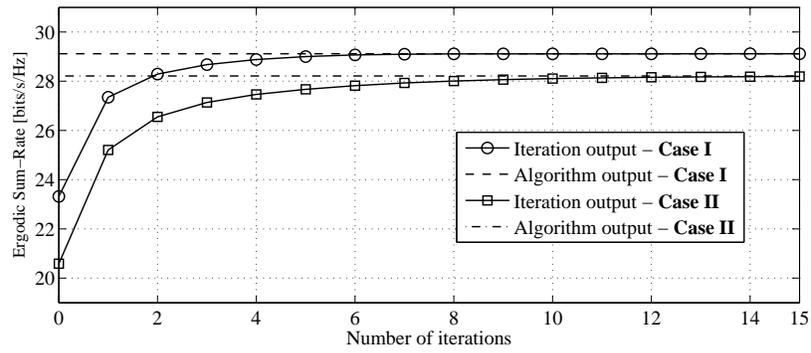}}}
   \label{fig8_2}
 }

\subfigure[Distributed algorithm]{
    \epsfxsize 5in \centerline{\hbox{\epsffile{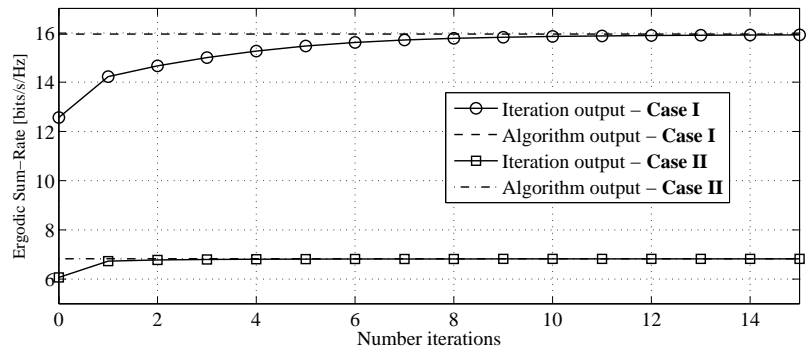}}}
   \label{fig8_3}
 }

\caption{Convergence behavior of three different proposed algorithms for $\textrm{SNR}=30$ dB.}\label{fig8}
\end{figure}

Finally, we present the convergence curves of the sum-rate for the proposed algorithms with respect to the number of iterations in Fig.~\ref{fig8}. For $\mathrm{SNR}=30$ dB, three proposed schemes in Fig. \ref{fig8_1}, \ref{fig8_2}, and \ref{fig8_3} were performed in the specific fading scenarios mentioned in section \ref{sec_design_slow}, \ref{sec_design_fast2}, and \ref{sec_design_fast1}, respectively. These results reveal that most of the proposed algorithms provide the sum-rate performance close to the outputs of the algorithms around 10 iterations, while the {\bf Case II} of the distributed algorithm shows very fast convergence behavior.


\section{Conclusion}
\label{sec_con}
We investigated in this paper a two-hop AF MIMO relaying network where three half-duplex relays help forward the message to the destination. An alternate relaying protocol and IA scheme were adopted to compensate for an inherent penalty of capacity pre-log factor $\frac{1}{2}$. The inter-relay interferences incurred by an alternate protocol were aligned to the reduced spatial dimensions and completely canceled at the relay. We aimed at optimizing source and relay filters to maximize the system achievable sum-rate and provided suboptimal solutions for different fading scenarios. Our proposed scheme can achieves $\frac{3M}{4}$ DOFs, while the conventional AF relaying schemes provide $\frac{M}{2}$ DOFs. From our simulation results, it was shown that the proposed filter designs are suitable for each fading scenario and have significant improvement over a naive filter, iterative IA scheme, and conventional half-duplex relaying schemes.

The generalization of the proposed system using arbitrary number of relays is our future work. Intuitively, as the number of relays increases, the achievable DOFs will increases. We will investigate the feasible strategy to transmit the achievable DOFs to the relays properly for each time slot.


%

\appendices
\section{Partial Derivative of $f_z(\mathbf{Z},\mathbf{Z}^*)=\ln\det\left(\mathbf{I}+p_t\mathbf{H}_z^\mathsf{H}\boldsymbol
\Sigma_z^{-1}\mathbf{H}_z\right)$}\label{app1}

First, we consider the MSE matrix which is unified as
\begin{eqnarray}
\mathbf{E}_z = \left(\mathbf{I}+p_t\mathbf{H}_z^\mathsf{H}\boldsymbol\Sigma_z^{-1}\mathbf{H}_z\right)^{-1},
\end{eqnarray}
where
$\mathbf{H}_z=\sum_{i=1}^l\sqrt{p_i}\mathbf{H}_{fi}\mathbf{X}_i\mathbf{H}_{bi}$,
$\boldsymbol\Sigma_z=\sum_{i=1}^l\sigma_i^2p_i\mathbf{H}_{fi}\mathbf{X}_i\mathbf{X}_i^\mathsf{H}\mathbf{H}_{fi}^\mathsf{H}+\sigma_D^2\mathbf{I}$,
and $p_i\mathrm{tr}\{\mathbf{X}_i\boldsymbol\Sigma_i\mathbf{X}_i^\mathsf{H}\} = P_R$
for any $p_t$, $P_R$, $\sigma_i^2$, $\sigma_D^2$, $\boldsymbol\Sigma_i$, $\mathbf{H}_{fi}$, and $\mathbf{H}_{bi}$. When $\mathbf{X}_i$ is a function of $\mathbf{Z}$, the differential of $f_z(\mathbf{Z},\mathbf{Z}^*)$ is computed as
$df_z(\mathbf{Z},\mathbf{Z}^*) = \mathrm{tr}\{\mathbf{E}_zd\mathbf{E}_z^{-1}\}$,
where we use the property, $\ln(\det(\mathbf{Z}))=\mathrm{tr}\{\mathbf{Z}^{\mathsf{-1}}d\mathbf{Z}\}$, in~\cite{Hj07}. The differential of $\mathbf{E}_z^{-1}$ is computed as
\begin{eqnarray}
d\mathbf{E}_z^{-1}\!&\!=\!&\!p_t\sum_{i=1}^l\left(d\sqrt{p_i}\mathbf{H}_{bi}^\mathsf{H}\mathbf{X}_i^\mathsf{H}
\mathbf{H}_{fi}^\mathsf{H}+\sqrt{p_i}\mathbf{H}_{bi}^\mathsf{H}d\mathbf{X}_i^\mathsf{H}\mathbf{H}_{fi}^\mathsf{H}\right)
\boldsymbol\Sigma_z^{-1}\mathbf{H}_z\nonumber\\
\!&\!+\!&\!p_t\sum_{i=1}^l\mathbf{H}_z^\mathsf{H}\boldsymbol\Sigma_z^{-1}\left(d\sqrt{p_i}\mathbf{H}_f\mathbf{X}_i
\mathbf{H}_b+
\sqrt{p_i}\mathbf{H}_fd\mathbf{X}_i\mathbf{H}_b\right),\label{dmse_Ez}\nonumber\\
\!&\!-\!&\!p_t\mathbf{H}_z^\mathsf{H}\boldsymbol\Sigma_z^{-1}d\boldsymbol\Sigma_z\boldsymbol\Sigma_z^{-1}
\mathbf{H}_z,
\end{eqnarray}
where
\begin{eqnarray}
dp_i\!&\!=\!&\!-\frac{p_i^2}{P_R}\mathrm{tr}\{d\mathbf{X}_i\boldsymbol\Sigma_i\mathbf{X}_i^\mathsf{H}
+\mathbf{X}_i\boldsymbol\Sigma_id\mathbf{X}_i^\mathsf{H}\},\nonumber\\
d\sqrt{p_i}\!&\!=\!&\!-\frac{p_i\sqrt{p_i}}{2P_R}\mathrm{tr}\{d\mathbf{X}_i\boldsymbol\Sigma_i
\mathbf{X}_i^\mathsf{H}
+\mathbf{X}_i\boldsymbol\Sigma_id\mathbf{X}_i^\mathsf{H}\},\nonumber\\
d\boldsymbol\Sigma_z^{-1}\!&\!=\!&\!\sum_{i=1}^l\sigma_i^2dp_i\mathbf{H}_f\mathbf{X}_i\mathbf{X}_i^\mathsf{H}
\mathbf{H}_f^\mathsf{H}+\sigma_i^2p_i\mathbf{H}_{fi}d\mathbf{X}_i\mathbf{X}_i^\mathsf{H}\mathbf{H}_{fi}^\mathsf{H}\nonumber.
\end{eqnarray}
Plugging (\ref{dmse_Ez}) into $\mathrm{tr}\{\mathbf{E}_zd\mathbf{E}_z^{-1}\}$ and applying some manipulation yields
\begin{eqnarray}
df_z(\mathbf{Z},\mathbf{Z}^*)=\mathrm{tr}\{\mathbf{E}_zd\mathbf{E}_z^{-1}\}\!&\!=\!&\!p_t\sum_{i=1}^l\mathrm{tr}\left\{\boldsymbol\Psi_{zi}
d\mathbf{X}_i^\mathsf{H}+\boldsymbol\Psi_{zi}^{\mathsf{H}}d\mathbf{X}_i\right\},\label{unified_trEz1}
\end{eqnarray}
In (\ref{unified_trEz1}), we define
\begin{eqnarray}
\boldsymbol\Psi_{zi}\!&\!=\!&\!\sqrt{p_i}\left[\boldsymbol\Omega_{zi}-\frac{p_i}{P_R}Re\left(\mathrm{tr}
\left\{\mathbf{X}_i^\mathsf{H}\boldsymbol\Omega_{zi}\right\}\right)\mathbf{X}_i\boldsymbol\Sigma_i\right],
\end{eqnarray}
where $\boldsymbol\Omega_{zi}=\mathbf{H}_{fi}^\mathsf{H}\boldsymbol\Sigma_z^{-1}\mathbf{H}_z\mathbf{E}_z
\left(\mathbf{H}_{bi}^\mathsf{H}-\sigma_i^2\sqrt{p_i}\mathbf{H}_z^\mathsf{H}\boldsymbol\Sigma_z^{-1}
\mathbf{H}_{fi}\mathbf{X}_i\right)$.
The differentials in (\ref{unified_trEz1}), $\mathbf{X}_i$ can be replaced with a function of $\mathbf{Z}$, and if $df_z(\mathbf{Z},\mathbf{Z}^*)=\mathrm{tr}\{\mathbf{A}_0^\mathsf{T}d\mathbf{Z}+\mathbf{A}_1^\mathsf{T}d\mathbf{Z}^\mathsf{*}\}$, it can be shown that $\partial f/\partial\mathbf{Z} = \mathbf{A}_0$ and $\partial f_z/\partial\mathbf{Z}^\mathsf{*} = \mathbf{A}_1$~\cite{Hj07}. Since we consider only partial derivative with respect to $\mathbf{Z}^*$ to utilize the subgradient method, we do not need to consider the differential with respect to $\mathbf{Z}$ and omit the first term for convenience, that is,
\begin{eqnarray}
df_z(\mathbf{Z}^*)=\mathrm{tr}\{\mathbf{A}_1^\mathsf{T}d\mathbf{Z}^\mathsf{*}\} \Rightarrow \frac{\partial f_z}{\partial\mathbf{Z}^\mathsf{*}} = \mathbf{A}_1.\label{rel_diff_par}
\end{eqnarray}

\subsection*{Example: Partial Derivatives of $f_1=\frac{1}{2}(I_o+I_e)$}
Let us consider $f_1=\frac{1}{2\ln2}(\ln\det\mathbf{E}_o^{-1}+\ln\det\mathbf{E}_e^{-1})$ in section \ref{sec_design_slow}. When $f_{z1}(\mathbf{Z},\mathbf{Z}^*)=\ln\det\mathbf{E}_o^{-1}$, we set $l=1$, $p_t=p_o$, $\mathbf{H}_{fi}=\mathbf{H}_{D3}$, $\mathbf{H}_{bi}=\mathbf{H}_{3S}\mathbf{T}_o$, and $\mathbf{X}_i=\mathbf{F}_3=\mathbf{U}_w^\perp\mathbf{G}_3\mathbf{U}_b^\perp$. Using these setting, we can compute
\begin{eqnarray}
df_{z1}(\mathbf{Z},\mathbf{Z}^*)=p_o\mathrm{tr}\left\{\boldsymbol\Psi_3d\mathbf{F}_3^\mathsf{H}+\boldsymbol\Psi_3^{\mathsf{H}}d\mathbf{F}_3\right\}.
\end{eqnarray}
For $\mathbf{Z}=\mathbf{U}_b$, the differential of $\mathbf{F}_3$ and $\mathbf{F}_3^\mathsf{H}$ with respect to $\mathbf{U}_b^\mathsf{H}$ can be computed as
$d\mathbf{F}_3=-\mathbf{U}_w^{\bot}\mathbf{G}_3\mathbf{U}_b^\dag d\mathbf{U}_b^\mathsf{H}\mathbf{U}_b^{\bot}$ and $d\mathbf{F}_3^\mathsf{H}=-\mathbf{U}_b^\dag d\mathbf{U}_b^\mathsf{H}\mathbf{U}_b^{\bot}\mathbf{G}_3^\mathsf{H}\mathbf{U}_w^{\bot}$. The differential of $f_{z1}(\mathbf{U}_b^*)$ can be presented as
\begin{eqnarray}
df_{z1}(\mathbf{U}_b^*)\!&\!=\!&\!-p_o\mathrm{tr}\left\{\left[\mathbf{U}_b^\perp\left(\mathbf{G}_3^\mathsf{H}\mathbf{U}_w^{\perp}\boldsymbol\Psi_3+\boldsymbol\Psi_3^{\mathsf{H}}
\mathbf{U}_w^{\perp}\mathbf{G}_3\right)\mathbf{U}_b^\dag\right]^\mathsf{T} d\mathbf{U}_b^*\right\}.
\end{eqnarray}
In the same way, we can find the differentials with respect to $\mathbf{U}_w$ and $\mathbf{G}_3$ which are given by
\begin{eqnarray}
df_{z1}(\mathbf{U}_w^*)\!&\!=\!&\!
-p_o\mathrm{tr}\left\{\left[\mathbf{U}_w^{\bot}\left(\boldsymbol\Psi_3\mathbf{U}_b^{\bot}\mathbf{G}_3^\mathsf{H}
+\mathbf{G}_3\mathbf{U}_b^{\bot}\boldsymbol\Psi_3^{\mathsf{H}}\right)\mathbf{U}_w^{\dag}\right]^\mathsf{T}d\mathbf{U}_w^*\right\},
\end{eqnarray}
and
\begin{eqnarray}
df_{z1}(\mathbf{G}_3^*)\!&\!=\!&\!\mathrm{tr}\left\{\left(\mathbf{U}_w^\perp\boldsymbol\Psi_3\mathbf{U}_b^\perp\right)^\mathsf{T} d\mathbf{G}_3^*\right\}.
\end{eqnarray}
Meanwhile, if $f_{z2}(\mathbf{Z},\mathbf{Z}^*)=\ln\det\mathbf{E}_e^{-1}$, we set $l=2$, $p_t=p_e$, $\mathbf{H}_{fi}=\mathbf{H}_{Di}$, $\mathbf{H}_{bi}=\mathbf{H}_{iS}\mathbf{T}_e$, and $\mathbf{X}_i=\mathbf{F}_i=\mathbf{H}_{i3}^{-1}\mathbf{U}_b\mathbf{G}_i\mathbf{U}_w^\mathsf{H}\mathbf{H}_{i3}^{-1}$. Plugging these values, the differential can be rewritten as
\begin{eqnarray}
df_{z2}(\mathbf{Z},\mathbf{Z}^*)=p_e\sum_{i=1}^2\mathrm{tr}\left\{\boldsymbol\Psi_id\mathbf{F}_i^\mathsf{H}+\boldsymbol\Psi_i^{\mathsf{H}}d\mathbf{F}_i\right\}.
\end{eqnarray}
For $\mathbf{Z}=\mathbf{U}_b$, we do not need to consider the second term for $d\mathbf{F}_i$ since we focus on only the differential with respect to $\mathbf{U}_b^*$. The differential of $\mathbf{F}_i^\mathsf{H}$ can be computed as
$d\mathbf{F}_i^\mathsf{H}=\mathbf{H}_{i3}^{-\mathsf{H}}\mathbf{U}_w\mathbf{G}_i^\mathsf{H}d\mathbf{U}_b^\mathsf{H}\mathbf{H}_{3i}^{-\mathsf{H}}$ and the differential of $f_{z2}(\mathbf{U}_b^*)$ can be represented as
\begin{eqnarray}
df_{z2}(\mathbf{U}_b^*)=p_e\sum_{i=1}^2\mathrm{tr}\left\{\left[\mathbf{H}_{3i}^{-\mathsf{H}}\boldsymbol\Psi_i\mathbf{H}_{i3}^{-\mathsf{H}}\mathbf{U}_w\mathbf{G}_i^\mathsf{H}\right]^\mathsf{T}
d\mathbf{U}_b^\mathsf{H}\right\}.
\end{eqnarray}
Similarly, for $\mathbf{Z}=\mathbf{U}_w$, discarding the first term for $d\mathbf{F}_i^\mathsf{H}$ and using $d\mathbf{F}_i=\mathbf{H}_{3i}^{-1}\mathbf{U}_b\mathbf{G}_id\mathbf{U}_w^\mathsf{H}\mathbf{H}_{i3}^{-1}$, the differential of $f_{z2}(\mathbf{U}_w^*)$ can be computed as
\begin{eqnarray}
df_{z2}(\mathbf{U}_w^*)=p_e\sum_{i=1}^2\mathrm{tr}\left\{\left[\mathbf{H}_{i3}^{-1}\boldsymbol\Psi_i^{\mathsf{H}}\mathbf{H}_{3i}^{-1}\mathbf{U}_b\mathbf{G}_i\right]^\mathsf{T}
d\mathbf{U}_w^\mathsf{H}\right\}.
\end{eqnarray}
In case of $\mathbf{Z}=\mathbf{G}_i$ for $i=1$, 2, we consider $d\mathbf{F}_i^\mathsf{H}=\mathbf{H}_{i3}^{-\mathsf{H}}\mathbf{U}_wd\mathbf{G}_i^\mathsf{H}\mathbf{U}_b^\mathsf{H}
\mathbf{H}_{3i}^{-\mathsf{H}}$ and the differential of $df_{z2}(\mathbf{G}_i^*)$ is given by
\begin{eqnarray}
df_{z2}(\mathbf{G}_i^*)=p_e\mathrm{tr}\left\{\left[\mathbf{U}_b^\mathsf{H}\mathbf{H}_{3i}^{-\mathsf{H}}
\boldsymbol\Psi_i\mathbf{H}_{i3}^{-\mathsf{H}}\mathbf{U}_w\right]d\mathbf{G}_i^\mathsf{H}\right\}.
\end{eqnarray}
Since $df_1(\mathbf{Z}^*)=\frac{1}{2\ln2}(df_z(\mathbf{Z}^*)+df_{z2}(\mathbf{Z}^*))$, we can obtain
\begin{eqnarray}
df_1(\mathbf{U}_b^*)\!&\!=\!&\!\frac{1}{2\ln2}\left(df_{z1}(\mathbf{U}_b^*)+df_{z2}(\mathbf{U}_b^*)\right)\nonumber\\
df_1(\mathbf{U}_w^*)\!&\!=\!&\!\frac{1}{2\ln2}\left(df_{z1}(\mathbf{U}_w^*)+df_{z2}(\mathbf{U}_w^*)\right)\nonumber\\
df_1(\mathbf{G}_i^*)\!&\!=\!&\!\frac{1}{2\ln2}df_{z2}(\mathbf{G}_i^*), \;\;i=1,2\nonumber\\
df_1(\mathbf{G}_3^*)\!&\!=\!&\!\frac{1}{2\ln2}df_{z1}(\mathbf{G}_3^*).\nonumber
\end{eqnarray}
Finally, using the relation in (\ref{rel_diff_par}), we can find the partial derivatives of $f_1$ with respect to $\mathbf{U}_b^*$, $\mathbf{U}_w^*$, $\mathbf{G}_1^*$, $\mathbf{U}_2^*$, and $\mathbf{G}_3^*$ given in (\ref{parUb}), (\ref{parUw}), (\ref{parG1andG2}), and (\ref{parG3}).

\section{Partial derivatives of $g_z(\mathbf{Z},\mathbf{Z}^*)=\ln\det\left(\mathbf{I}+\sum_{i=1}^l\frac{p_t}{\sigma_i^2}\mathbf{Y}^\mathsf{H}
\mathbf{H}_{bi}^\mathsf{H}\mathbf{X}_i^\dag\mathbf{X}_i^\mathsf{H}\mathbf{H}_{bi}\mathbf{Y}\right)$}\label{app2}

Now we consider the MSE matrix, $\mathbf{E}_z=\left(\mathbf{I}+\sum_{i=1}^l\frac{p_t}{\sigma_i^2}\mathbf{Y}^\mathsf{H}
\mathbf{H}_{bi}^\mathsf{H}\mathbf{X}_i^\dag\mathbf{X}_i^\mathsf{H}\mathbf{H}_{bi}\mathbf{Y}\right)^{-1}$, where $\mathbf{X}_i$ or $\mathbf{Y}$ is a function of $\mathbf{Z}$ and $p_t\mathrm{tr}\{\mathbf{Y}\mathbf{Y}^\mathsf{H}\}=P_S$. First of all, we find the differential of $g_z(\mathbf{Z},\mathbf{Z}^*)$ when $\mathbf{X}_i$ is a function of $\mathbf{Z}$.
The differential of $g_z(\mathbf{Z},\mathbf{Z}^*)$ is computed as $dg_z(\mathbf{Z},\mathbf{Z}^*)=\mathrm{tr}\{\mathbf{E}_zd\mathbf{E}_z^{-1}\}$, where
\begin{eqnarray}
d\mathbf{E}_z^{-1}=\sum_{i=1}^l\frac{p_t}{\sigma_i^2}\mathbf{Y}^\mathsf{H}\mathbf{H}_{bi}^\mathsf{H}\left(\mathbf{X}_i^\dag d\mathbf{X}_i^\mathsf{H}\mathbf{X}_i^\perp+\mathbf{X}_i^\perp
d\mathbf{X}_i\mathbf{X}_i^{\dag\mathsf{H}}\right)\mathbf{H}_{bi}\mathbf{Y}.
\end{eqnarray}
After some manipulation, we can obtain
\begin{eqnarray}
dg_z(\mathbf{Z},\mathbf{Z}^*)=\sum_{i=1}^l\frac{p_t}{\sigma_i^2}\mathrm{tr}\left\{\mathbf{X}_i^\perp\boldsymbol\Upsilon_{zi}
\mathbf{X}_i^\dag d\mathbf{X}_i^\mathsf{H}+\mathbf{X}_i^{\dag\mathsf{H}}
\boldsymbol\Upsilon_{zi}\mathbf{X}_i^\perp d\mathbf{X}_i\right\},\label{dgz_x}
\end{eqnarray}
where we define $\boldsymbol\Upsilon_{zi}=\mathbf{H}_{bi}\mathbf{Y}\mathbf{E}_z\mathbf{Y}^\mathsf{H}\mathbf{H}_{bi}^\mathsf{H}$.

Secondly, when $\mathbf{Z}$ is a function of $\mathbf{Y}$, the differential of $\mathbf{E}_z^{-1}$ is computed as
\begin{eqnarray}
d\mathbf{E}_z^{-1}=\sum_{i=1}^l\frac{dp_t}{\sigma_i^2}\mathbf{Y}^\mathsf{H}\mathbf{H}_{bi}^\mathsf{H}\mathbf{X}_i^\dag\mathbf{X}_i^\mathsf{H}\mathbf{H}_{bi}\mathbf{Y}+
\frac{p_t}{\sigma_i^2}d\mathbf{Y}^\mathsf{H}\mathbf{H}_{bi}^\mathsf{H}\mathbf{X}_i^\dag\mathbf{X}_i^\mathsf{H}\mathbf{H}_{bi}\mathbf{Y}+
\frac{p_t}{\sigma_i^2}\mathbf{Y}^\mathsf{H}\mathbf{H}_{bi}^\mathsf{H}\mathbf{X}_i^\dag\mathbf{X}_i^\mathsf{H}\mathbf{H}_{bi}d\mathbf{Y},
\end{eqnarray}
where $dp_t=-\frac{p_t^2}{P_S}\mathrm{tr}\{d\mathbf{Y}\mathbf{Y}^\mathsf{H}+\mathbf{Y}d\mathbf{Y}^\mathsf{H}\}$.
Plugging it into $\mathrm{tr}\{\mathbf{E}_zd\mathbf{E}_z^{-1}\}$, the differential of $g_z(\mathbf{Z},\mathbf{Z}^*)$ is computed as
\begin{eqnarray}
dg_z(\mathbf{Z},\mathbf{Z}^*)\!&\!=\!&\!\sum_{i=1}^l\mathrm{tr}\left\{\frac{p_t}{\sigma_i^2}\mathbf{H}_{bi}^\mathsf{H}\mathbf{X}_i^\dag\mathbf{X}_i^\mathsf{H}\mathbf{H}_{bi}\mathbf{Y}\mathbf{E}_z
d\mathbf{Y}^\mathsf{H}-\frac{p_t^2}{\sigma_i^2P_S}\mathrm{tr}\{\boldsymbol\Upsilon_{zi}\mathbf{X}_i^\dag\mathbf{X}_i^\mathsf{H}\}\mathbf{Y}d\mathbf{Y}^\mathsf{H}\right\}\nonumber\\
\!&\!+\!&\!\sum_{i=1}^l\mathrm{tr}\left\{\frac{p_t}{\sigma_i^2}\mathbf{E}_z\mathbf{Y}^\mathsf{H}\mathbf{H}_{bi}^\mathsf{H}\mathbf{X}_i^\dag\mathbf{X}_i^\mathsf{H}\mathbf{H}_{bi}
d\mathbf{Y}-\frac{p_t^2}{\sigma_i^2P_S}\mathrm{tr}\{\boldsymbol\Upsilon_{zi}\mathbf{X}_i^\dag\mathbf{X}_i^\mathsf{H}\}\mathbf{Y}^\mathsf{H}d\mathbf{Y}\right\}.\label{dgz_y}
\end{eqnarray}
As introduced in (\ref{rel_diff_par}) of Appendix~\ref{app1}, we can find the partial derivative with respect to $\mathbf{Z}^*$.

\subsection*{Example: Partial Derivatives of $f_2=\frac{1}{2}(I_o+I_c)$}
When we consider $f_2=\frac{1}{2\ln2}(\ln\det\mathbf{E}_o^{-1}+\ln\det\mathbf{E}_c^{-1})$ in section \ref{sec_design_fast}. When we define $g_{z1}(\mathbf{Z},\mathbf{Z}^*)=\ln \det\mathbf{E}_c^{-1}$, $f_2$ can be rewritten as $f_2(\mathbf{Z},\mathbf{Z}^*)=\frac{1}{2\ln2}\left(f_{z1}(\mathbf{Z},\mathbf{Z}^*)+g_{z1}(\mathbf{Z},\mathbf{Z}^*)\right)$, where we set $l=2$, $p_t=p_e$, $\mathbf{H}_{bi}=\mathbf{H}_{iS}$, $\mathbf{X}_i=\mathbf{W}_i$, and $\mathbf{Y}=\mathbf{T}_e$ for $g_{z1}(\mathbf{Z},\mathbf{Z}^*)$. The differential of $f_2$ is computed as
\begin{eqnarray}
df_2(\mathbf{Z},\mathbf{Z}^*)=\frac{1}{2\ln2}\left(df_{z1}(\mathbf{Z},\mathbf{Z}^*)+dg_{z1}(\mathbf{Z},\mathbf{Z}^*)\right).
\end{eqnarray}
$df_{z1}(\mathbf{Z},\mathbf{Z}^*)$ has been considered in Appendix~\ref{app1} and we focus on $dg_{z1}(\mathbf{Z},\mathbf{Z}^*)$. First, using (\ref{dgz_x}), the differential of $g_{z1}(\mathbf{Z},\mathbf{Z}^*)$ can be represented as
\begin{eqnarray}
dg_{z1}(\mathbf{Z},\mathbf{Z}^*)=\sum_{i=1}^2\frac{p_e}{\sigma_i^2}\mathrm{tr}\left\{\mathbf{W}_i^{\bot}
\boldsymbol\Upsilon_i\mathbf{W}_i^\dag d\mathbf{W}_i^\mathsf{H}+\mathbf{W}_i^{\dag\mathsf{H}}
\boldsymbol\Upsilon_i\mathbf{X}_i^{\bot}d\mathbf{W}_i\right\}.
\end{eqnarray}
For $\mathbf{Z}=\mathbf{U}_w$, where $\mathbf{W}_i=\mathbf{H}_{i3}^{-\mathsf{H}}\mathbf{U}_w$, we do not need to consider the second term for $d\mathbf{W}_i$ since we should find the differential with respect to $\mathbf{U}_w^*$. Substituting $d\mathbf{W}_i^\mathsf{H}=d\mathbf{U}_w^\mathsf{H}\mathbf{H}_{i3}^{-1}$ yields
\begin{eqnarray}
dg_{z1}(\mathbf{U}_w^*)=\sum_{i=1}^2\frac{p_e}{\sigma_i^2}\mathrm{tr}\left\{\mathbf{H}_{i3}^{-1}
\mathbf{W}_i^{\bot}\boldsymbol\Upsilon_1\mathbf{W}_i^\dag d\mathbf{U}_w^\mathsf{H}\right\}.
\end{eqnarray}
Secondly, using (\ref{dgz_y}) for $\mathbf{Z}=\mathbf{T}_e$, $dg_{z1}(\mathbf{T}_e^*)$ can be computed as
\begin{eqnarray}
dg_{z1}(\mathbf{T}_e^*)=\sum_{i=1}^2\mathrm{tr}\left\{\frac{p_e}{\sigma_i^2}\mathbf{H}_{iS}^\mathsf{H}
\mathbf{W}_i^\dag\mathbf{W}_i^\mathsf{H}\mathbf{H}_{iS}\mathbf{T}_e\mathbf{E}_cd\mathbf{T}_e^\mathsf{H}
-\frac{p_e^2}{\sigma_i^2P_S}\mathrm{tr}\{\boldsymbol\Upsilon_i\mathbf{W}_i^\dag\mathbf{W}_i^\mathsf{H}
\}\mathbf{T}_e^\mathsf{H}d\mathbf{T}_e^\mathsf{H}\right\}.
\end{eqnarray}
Finally, the differential of $f_2$ with respect to $\mathbf{U}_w^*$ and $\mathbf{T}_e^*$ can be computed as
\begin{eqnarray}
df_2(\mathbf{U}_w^*)\!&\!=\!&\!\frac{1}{2\ln2}(df_{z1}(\mathbf{U}_w^*)+dg_{z1}(\mathbf{U}_w^*))\nonumber\\
df_2(\mathbf{T}_e^*)\!&\!=\!&\!\frac{1}{2\ln2}dg_{z1}(\mathbf{T}_e^*).\nonumber
\end{eqnarray}


\ifCLASSOPTIONcaptionsoff
  \newpage
\fi



%
%
%

\bibliographystyle{ieeetran}
\bibliography{IEEEabrv,mybib}

%
%
%
%
%



\newpage

\begin{table}
\begin{center}
\caption{Protocol} \label{protocol}\vspace{0.2in}
\begin{tabular}{|c||c|c|}\hline
Transmission Type / Time Slot & even time slot & odd time slot  \\ \hline
 Source Transmission & $S\longrightarrow R_1,R_2$ & $S\longrightarrow R_3$   \\ \hline
 Relaying & $R_3\longrightarrow D$ & $R_1,R_2\longrightarrow D$   \\ \hline
\end{tabular}
\end{center}
\end{table}

\clearpage
\newpage

\clearpage
\newpage

\clearpage
\newpage

\clearpage
\newpage

\clearpage
\newpage

\clearpage
\newpage

\clearpage

\newpage

\end{document}